\documentclass[oldversion]{aa}

\usepackage[tbtags,fleqn]{amsmath}
\usepackage{txfonts}
\usepackage{amsfonts}
\usepackage{pstricks, pst-plot}
\usepackage{graphicx}
\usepackage{natbib}

\newcommand{\diff}{\mathrm d}

\newcommand{\mincir}{\raise
  -2.truept\hbox{\rlap{\hbox{$\sim$}}\raise5.truept \hbox{$<$}\ }}
\newcommand{\magcir}{\raise
  -2.truept\hbox{\rlap{\hbox{$\sim$}}\raise5.truept \hbox{$>$}\ }}


\allowdisplaybreaks[1]

\begin{document}

\title{\textsc{Nicest}, a near-infrared color excess method tailored
  for small-scale structures.}  \titlerunning{\textsc{Nicest}}
\author{Marco Lombardi\inst{1,2}} \offprints{M. Lombardi}
\mail{mlombard@eso.org} \institute{%
  European Southern Observatory, Karl-Schwarzschild-Stra\ss e 2,
  D-85748 Garching bei M\"unchen, Germany \and University of Milan,
  Department of Physics, via Celoria 16, I-20133 Milan, Italy}
\date{Received ***date***; Accepted ***date***} \abstract{%
  Observational data and theoretical calculations show that
  significant small-scale substructures are present in dark molecular
  clouds.  These inhomogeneities can provide precious hints on the
  physical conditions inside the clouds, but can also severely bias
  extinction measurements.  We present \textsc{Nicest}, a novel method
  to account and correct for inhomogeneities in molecular cloud
  extinction studies.  The method, tested against numerical
  simulations, removes almost completely the biases introduced by
  sub-pixel structures and by the contamination of foreground stars.
  We applied \textsc{Nicest} to 2MASS data of the Pipe molecular
  complex.  The map thereby obtained shows significantly higher (up to
  $0.41 \mbox{ mag}$ in $A_K$) extinction peaks than the standard
  \textsc{Nicer} (Lombardi et al.\ 2001) map.  This first application
  confirms that substructures in nearby molecular clouds, if not
  accounted for, can significantly bias extinction measurements in
  regions with $A_K > 1 \mbox{ mag}$; the effect, moreover, is
  expected to increase in more distant molecular cloud, because of the
  poorer physical resolution achievable.  \keywords{dust, extinction
    -- methods: statistical -- ISM: clouds -- infrared: ISM -- ISM:
    structure -- ISM: individual objects: Pipe molecular complex}%
}

\maketitle

\section{Introduction}
\label{sec:introduction}

Although the details of star and planet formation are very little
known, there is a large consensus that these objects are created
inside dark molecular clouds from the contraction and fragmentation of
dense, cold cores.  As a result, in the last decades molecular clouds
have been studied in detail using many different techniques, from
optical number counts \citep{1923AN....219..109W,
  1937dss..book.....B}, to radio observations of carbon monoxide (CO)
molecules \citep{1970ApJ...161L..43W,1982ApJ...262..590F}, to
near-infrared (NIR) extinction measurements
\citep{1994ApJ...429..694L}.

Molecular hydrogen and helium represent approximately $99\%$ of the
mass of a cloud, but the absence of a dipole moment in these molecules
makes them virtually undetectable at the low temperatures ($T \sim 10
\mbox{ K}$) present in these objects.  Hence, the (projected) mass
distribution of dark clouds is usually inferred from the distribution
of relatively rare tracers (such as CO or dust grains) under the
assumption that these are uniformly distributed in the cloud.

As pointed out by \citet{1994ApJ...429..694L}, the reddening of
background stars in NIR bands is a simple and reliable method to study
the dust distribution and thus the hydrogen column density.  Compared
to optical bands, NIR bands are less affected by extinction and are
less sensitive to the physical properties of the dust grains
\citep{1990ARA&A..28...37M}, and thus their color excess can be
directly translated into a hydrogen column density.  In a series of
papers we reconsidered and improved the NIR color excess
(\textsc{Nice}) technique, by generalizing it to use three or more
bands (\textsc{Nicer}, see \citealp{2001A&A...377.1023L}) and by
taking a maximum-likelihood approach \citep{2005A&A...438..169L}.

Although the \textsc{Nice} and \textsc{Nicer} techniques have been
very successful in studying molecular clouds \citep[see,
e.g.,][]{2001Natur.409..159A}, they are plagued by two
complications: small-scale inhomogeneities and contamination by
foreground stars.  NIR studies typically assume a uniform extinction
over (small) patches of the sky; however, in presence of substructures
such as steep gradients, unresolved filaments, or turbulence-induced
inhomogeneities \citep{1994ApJ...429..694L, 1981MNRAS.194..809L,
  2004ApJ...615L..45H}, the background stars used to estimate the
cloud extinction would not be uniformly distributed in the patch, but
would be preferentially observed in low density regions.  Similarly,
in presence of foreground stars, their \textit{relative\/} fraction
increases with the dust column density.  Unfortunately, both effects
will bias \textit{all\/} color excess estimators toward low column
densities; moreover, the bias will be especially important in the
dense regions of molecular clouds, which is where the star formation
takes place.

In this paper we describe in details the repercussions of small scale
inhomogeneities and foreground stars in extinction measurements, and
propose a simple method to obtain estimates of the column density in
presence of both effects that, under simple working hypotheses, is
asymptotically unbiased.  Our method does not rely on any (usually
poorly verified) assumption regarding the small scale structure of the
cloud; moreover, it can be applied to a general class of NIR
extinction measurements (marked spatial point processes).

The paper is organized as follows.  In
Sect.~\ref{sec:cloud-substructure} we introduce a general smoothing
technique virtually used in all cases to create smooth, continuous
maps from the discrete pencil-beam extinction measurements carried out
for each background star; we also show that in two typical
applications (moving average and nearest neighbours smoothing) a bias
is expected.  In Sect.~\ref{sec:over-weighting-high} we propose a new
method that can correct this bias without making any assumption on the
underlying form of the cloud substructure.  This new method has been
tested with numerical simulations, as described in
Sect.~\ref{sec:simulations}, and with a real-case application,
presented in Sect.~\ref{sec:sample-application}.  We discuss and
comment the results obtained in Sect.~\ref{sec:discussion}, and
finally we briefly present our conclusions in
Sect.~\ref{sec:conclusions}.

\section{Cloud substructure}
\label{sec:cloud-substructure}

Measurements of the reddening of stars observed through a molecular
cloud provide estimates of the cloud column densities on pencil beams
characterized by a diameter of the order of a fraction of milliarcsec.
However, these high resolution measurements are strongly undersampled
and are affected by large uncertainties due to the photometric errors
and to the intrinsic scatter of star colors in the NIR.  Smooth and
continuous extinction maps are normally obtained by interpolating and
binning the various pencil beams.  Different authors use different
recipes to smooth the data (see, e.g., \citealp{2002A&A...395..733L}
for a description of several interpolation techniques), but in most
cases the interpolation follows a standard scheme.  Consider $N$
$K$-band extinction\footnote{For simplicity, in this paper we drop the
  subscript $K$ normally used in literature to denote the $K$-band
  extinction $A_K$; it is also obvious that the same method applies to
  extinction measurements referred to any band (e.g., the visual
  extinction $A_V$).} measurements $\bigl\{ \hat A_n \bigr\}$
obtained, for example, from the color excess of $N$ background (see
\citealp{1994ApJ...429..694L} or \citealp{2001A&A...377.1023L}).  The
extinction $\hat A$ at any location in the sky is evaluated using a
weighted average
\begin{equation}
  \label{eq:1}
  \hat A = \frac{\sum_n w_n \hat A_n}{\sum_n w_n} \; .
\end{equation}
The weights $\{ w_n \}$ are usually chosen to be significantly
different from zero only for stars angularly close to the given
location of the map (for example, \citealp{2002AJ....123.2559C} assign
a unity weight to the $N$ nearest neighbors).  Clearly,
Eq.~\eqref{eq:1} does not include slightly more complex situations
where, for example, one takes the median of the extinctions measured
from objects nearby a given position
\citep[e.g.][]{2008A&A...484..205D}; however, most of the discussion
carried out for the weighted average actually applies also to median
or related estimators.  

The binning in Eq.~\eqref{eq:1} washes out the cloud substructure on
scales smaller than the typical size where the $\{ w_n \}$ are
significantly different from zero, but this is needed in order to have
smooth maps and to increase the signal-to-noise ratio.  However,
Eq.~\eqref{eq:1} also introduces a significant bias on the estimated
column density.  Suppose that, in the region of the sky that we are
investigating (i.e., in the area covered by the $N$ stars used to
estimate $A$) the column density has significant variations.  Because
of this, the local density $\rho(\vec x)$ of stars is not homogeneous
though the cloud, but rather follows the scheme
\begin{equation}
  \label{eq:2}
  \rho(\vec x) = \rho_0 10^{-\alpha k_\lambda A(\vec x)} \; ,
\end{equation}
where $\rho_0$ is the density of stars where no extinction is present,
$\alpha$ is the slope of the number counts, and $k_\lambda = A_\lambda
/ A$ is the extinction law in the band $\lambda$ considered.  This
effect, which is at the origin of the historical number count method
to measure column densities in molecular clouds
\citep{1923AN....219..109W, 1937dss..book.....B}, also induces a bias
in Eq.~\eqref{eq:1}, since regions with smaller extinctions and thus
higher density of background stars will, on average, contribute more
to the sum in Eq.~\eqref{eq:1} than regions with large extinctions.
Note that since the color excess method requires measurements in at
least two different bands, one should replace Eq.~\eqref{eq:2} with a
version that provides the expected density of stars observed in all
bands required for the application of the method (e.g., $H$ and $K$).
For equally deep observations on all bands, the result has the same
form of Eq.~\eqref{eq:2}, where $k_\lambda$ is the extinction law of
the bluer band considered (e.g., $H$).

The main issue considered in this paper is the bias introduced by
unresolved substructures in Eq.~\eqref{eq:1}.  In general, the bias
should be defined here as the difference between the expected, mean
value of $\langle \hat A \rangle$ and the true column density $A$ at
the same position.  However, clearly every smoothing technique
introduces a bias because of the smoothing itself, and this bias is
normally acceptable if it does not introduces a systematic effect on
the \textit{total\/} column density.  In other words, a required
property is
\begin{equation}
  \label{eq:3}
  \int A(\vec x) \, \diff x = \left\langle \int \hat A(\vec x) \, \diff x
  \right\rangle = \int \bigl\langle \hat A(\vec x) \bigr\rangle \, \diff x
  \; .
\end{equation}
Clearly, in order for Eq.~\eqref{eq:3} to hold, the difference
$\langle \hat A \rangle - A$ must be of alternating sign.  For our
purposes, thus, it is more useful to define the bias as the difference
between the expected mean value $\langle \hat A \rangle$ and the same
quantity that we would theoretically expect \textit{in presence of a
  uniform density of stars}, i.e.\ by ignoring the dependence of the
density of stars on the local extinction described by Eq.~\eqref{eq:2}
[see also below Eq.~\eqref{eq:7}].  This definition is useful, since
it isolates the standard effects of a smoothing from the effects
introduced by the non-uniform sampling of the cloud structure.  In
addition, a column density estimator $\hat A(\vec x)$ that is unbiased
according to this definition will also be unbiased according to
Eq.~\eqref{eq:3} (provided the weights $w_n$ are spatially invariant).
For, when the density of background stars is uniform, $\bigl\langle
\hat A(\vec x) \bigr\rangle$ can always be written as a convolution of
the true column density map $A(\vec x)$ with some kernel $K$
normalized to unity \citep{2002A&A...395..733L}.

In the following, we will calculate explicitly this bias in two
common smoothing schemes.

\subsection{Moving average smoothing}
\label{sec:moving-aver-smooth}

In this section we will consider a simple weighting scheme in
Eq.~\eqref{eq:1} where each weight $w_n = w(\vec x_n; \vec x)$ is a
simple function of the location $\vec x_n$ of the $n$-th star
(typically, $w(\vec x_n; \vec x_n)$ will depend only on the modulus
$\lvert \vec x_n - \vec x \rvert$).

As discussed above, our aim is to compare the expected column density
estimate with the one that we would obtain in absence of any selection
effect in the number of background stars.  We need thus to evaluate
two averages: (i) $\bigl\langle \hat A(\vec x) \bigr\rangle$, where $\hat
A$ is calculated according to Eq.~\eqref{eq:1} and the ensemble
average is taken over all positions of stars with the density given by
Eq.~\eqref{eq:2} and over all individual column density measurements
$\hat A_n$; and (ii) the average $\bar A(\vec x)$, which is basically the
same quantity evaluated with a constant density of background stars
$\rho_0$.

In principle, both averages can be calculated analytically using the
method described by \citeauthor{2001A&A...373..359L}
(\citeyear{2001A&A...373..359L}; see below
Sect.~\ref{sec:over-weighting-high}); in practice, for the purposes of
this section a simple approximation can be used provided that in the
weighted average of Eq.~\eqref{eq:1} a relatively large number of
column densities are used with weights significantly different from 0.
In this case we find for the first average
\begin{equation}
  \label{eq:4}
  \bigl\langle \hat A(\vec x) \bigr\rangle = \frac{\int A(\vec x') w(\vec
    x'; \vec x) \rho(\vec x') \, \diff x'}{\int w(\vec x'; \vec x)
    \rho(\vec x') \, \diff x'} \; .
\end{equation}
As shown by Eq.~\eqref{eq:2}, $\rho(\vec x')$ is a simple function of
$A(\vec x')$.  This suggests that the equation can be recast in a
simpler form by defining $p_A(A; \vec x)$, the (weighted) probability
that a point with extinction $A$ is used to evaluate $\hat A(\vec x)$,
the column density at $\vec x$:
\begin{equation}
  \label{eq:5}
  p_A(A; \vec x) = \frac{\int w(\vec x'; \vec x) \delta\bigl( A -
    A(\vec x') \bigr) \, \diff x'}{\int w(\vec x'; \vec x) \, \diff x'} \; .
\end{equation}
By using Eq.~\eqref{eq:5} in Eq.~\eqref{eq:4} we find then
\begin{equation}
  \label{eq:6}
  \bigl\langle \hat A(\vec x) \bigr\rangle = \frac{\int A p_A(A; \vec
    x) \rho(A) \, \diff A}{\int p_A(A; \vec x) \rho(A) \, \diff A} =
  \frac{\int A p_A(A; \vec x) 10^{-\alpha k_\lambda A} \, \diff
    A}{\int p_A(A; \vec x) 10^{-\alpha k_\lambda A} \, \diff A} \; .
\end{equation}
In presence of a uniform distribution of stars, $\rho(\vec x) =
\rho_0$, we would instead obtain simply
\begin{equation}
  \label{eq:7}
  \bar A(\vec x) = \frac{\int A(\vec x') w(\vec x'; \vec x) \, \diff
    x'}{\int w(\vec x'; \vec x) \, \diff x'} = \int A p_A(A; \vec x)
  \, \diff A \; . 
\end{equation}
The two values presented in Eqs.~\eqref{eq:6} and \eqref{eq:7} do not
differ significantly provided the scatter of $A$ in the patch
considered around $\vec x$ is much smaller than $(\alpha k_\lambda \ln
10)^{-1}$. For example, if we consider the $H$ band of a typical
region close to the Galactic plane with a standard reddening law
\citep{1985ApJ...288..618R}, we have $\alpha \simeq 0.34 \mbox{
  mag}^{-1}$ and $k_H \simeq 1.55$ \citep{2005ApJ...619..931I}, so
that the maximum scatter allowed in $A_K$ to have a negligible bias
is${} \ll 0.82 \mbox{ mag}$.  Hence, clearly we need to be concerned
by this bias in regions from moderate to large extinction.

It is also interesting to evaluate the bias $\bigl\langle \hat A(\vec
x) \bigr\rangle - \bar A(\vec x)$ in presence of small variations of
$A$ within the region considered.  In this case, the probability
distribution $p_A(A; \vec x)$ is peaked around the mean value of
Eq.~\eqref{eq:7}, and we can expand to second order the two
exponential functions present in the numerator and denominator of
Eq.~\eqref{eq:6}.  After a few manipulations we obtain then
\begin{align}
  \label{eq:8}
  \bigl\langle \hat A(\vec x) \bigr\rangle - \bar A(\vec x) \simeq {}
  & -\alpha k_\lambda \ln 10 \int p_A(A; \vec x) \bigl[ A - \bar
  A(\vec x) \bigr]^2 \, \diff A \notag\\
  {} = {} & - \beta \frac{\int w(\vec x'; \vec x) \Delta^2(\vec x';
    \vec x) \, \diff x'}{\int w(\vec x'; \vec x) \, \diff x'} \; ,
\end{align}
where $\beta \equiv \alpha k_\lambda \ln 10$ and $\Delta(\vec x'; \vec
x) \equiv A(\vec x') - \bar A(\vec x)$.  Hence, the difference between
the two values is proportional to a weighted average of the scatter of
$A$ around its mean value $\bar A(\vec x)$ at $\vec x$, a quantity
that, as we will see below, can be directly estimated from the data.
Note that the bias is, to first order, quadratic on $\Delta(\vec x';
\vec x)$ and thus will be particularly severe when steep gradients are
present in the underlying column density $A(\vec x)$.

\subsection{Nearest neighbor(s)}
\label{sec:nearest-neighbors}

\citet{2002AJ....123.2559C} suggest to use a different prescription to
make smooth extinction map from the individual column density
measurements.  For each point in the map, they take the average
extinction of the $N$ angularly closest stars (nearest neighbours
interpolation).  As argued by \citet{2005A&A...435..131C,
  2006A&A...445..999C}, this method can potentially alleviate the
bias introduced by the varying background density of stars described
by Eq.~\eqref{eq:2}, because measurements from low density regions
will mostly appear isolated and will thus be used for relatively large
patches of the sky.

Interestingly, using the technique described by
\citet{2002A&A...395..733L} it is possible to obtain an analytical
estimate for the average of $\hat A$ in the smoothing scheme proposed:
\begin{equation}
  \label{eq:9}
  \bigl\langle \hat A(\vec x) \bigr\rangle = \int A(\vec x') K(\vec x';
  \vec x) \rho(\vec x') \, \diff x' \; ,
\end{equation}
where the linear kernel $K(\vec x'; \vec x)$ is given by
\begin{equation}
  \label{eq:10}
  K(\vec x'; \vec x) = \frac{\mathrm{e}^{-\mu(\vec x'; \vec x)}}{N}
  \sum_{n=0}^{N-1} \frac{\bigl[ \mu(\vec x'; \vec x)
    \bigr]^n}{n!} \; .
\end{equation}
In this equation, $N$ is the number of stars used in the nearest
neighbors interpolation and $\mu(\vec x'; \vec x)$ is the average
number of stars observed in $B(\vec x'; \vec x)$, the disk centered in
$\vec x$ and of angular radius $\lvert \vec x - \vec x' \rvert$:
\begin{equation}
  \label{eq:11}
  \mu(\vec x'; \vec x) = \int_{B(\vec x'; \vec x)} \rho(\vec
  x'') \diff x'' \; .
\end{equation}
As shown by \citet{2002A&A...395..733L}, the kernel $K(\vec x'; \vec
x)$ is normalized, i.e.\
\begin{equation}
  \label{eq:12}
  \int K(\vec x'; \vec x) \rho(\vec x') \, \diff x' = 1 \; ,
\end{equation}
an obvious result if one consider the measurement of the extinction on
a uniform cloud where $A(\vec x) = \mbox{const}$.  In our case, this
property can also be proved directly from Eqs.~\eqref{eq:10} and
\eqref{eq:11}.  First, note that $\mu(\vec x'; \vec x) = \mu(r; \vec
x)$ depends only on $\vec x$ and on $r = \lvert \vec x - \vec x'
\rvert$, and thus the same applies to $K(\vec x'; \vec x) = K(r; \vec
x)$.  This suggests that we can recast the integral of
Eq.~\eqref{eq:12} as
\begin{multline}
  \label{eq:13}
  \int_0^\infty K(r; \vec x) \frac{\partial \mu(r; \vec x)}{\partial
    r} \, \diff r \\ = \sum_{n=0}^{N-1} \frac{n!}{N} \int_0^\infty
  \mathrm{e}^{-\mu(r; \vec x)} \bigl[ \mu(r; \vec x) \bigr]^n
  \frac{\partial \mu(r; \vec x)}{\partial r} \, \diff r = 1 \; .
\end{multline}

\begin{figure}
  \centering
  \includegraphics[width=\hsize, bb=28 9 408 264]{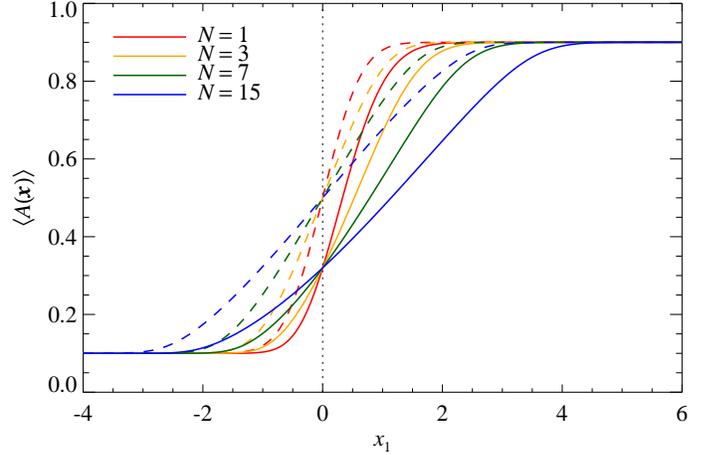}
  \caption{The bias of the nearest neighbors interpolation on a toy
    model where the true extinction is $A(\vec x) = 0.1 \mbox{ mag}$
    for $x_1 < 0$ and $0.9 \mbox{ mag}$ for $x_1 \ge 0$.  The solid
    lines show the average measured extinction of the nearest
    neighbours method, calculated using Eq.~\eqref{eq:9} for various
    number of neighbours $N$.  The dashed lines show the extinction
    that one would measure if the density of stars were uniform on the
    whole field.  Note that the solid lines are always below the
    dashed lines except at the extremes of this plot (where all
    estimators agree with the true value of $A$ there).  For this plot
    we set $\alpha = 0.34$, $k = 1.55$, and $\rho_0 = 1$.}
  \label{fig:1}
\end{figure}

\begin{figure}
  \centering
  \includegraphics[width=\hsize, bb=28 9 408 264]{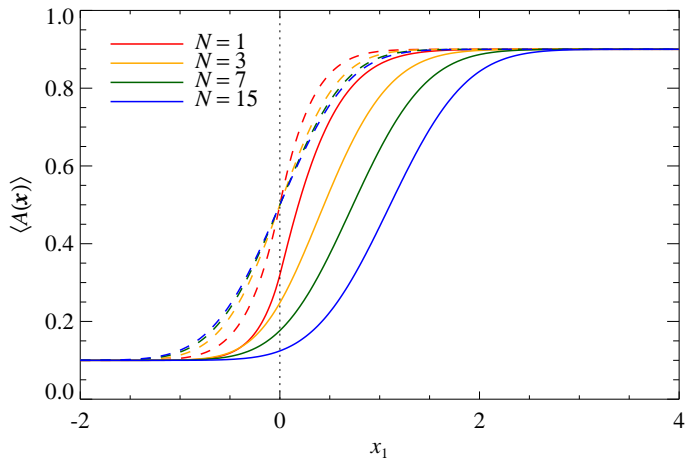}
  \caption{Similarly to Fig.~\ref{fig:1}, but for the median-nearest
    neighbors interpolation.  Again, the solid lines, representing the
    expected mean measurements, are always below the dashed lines,
    representing the same measurements in the ideal case where the
    density of background stars is uniform within the whole field.
    Note that because of the properties of the median, for large
    values of $N$ the ``transition'' from $A = 0.1 \mbox{ mag}$ to $A
    = 0.9 \mbox{ mag}$ is faster than in Fig.~\ref{fig:1}.}
  \label{fig:2}
\end{figure}

The fact that $K(\vec x'; \vec x)$ only depends on $\vec x$ and $r =
\lvert \vec x - \vec x' \rvert$ (and not on the direction of the
vector $\vec x - \vec x'$) also implies that the nearest neighbors
interpolation suffers from the bias discussed in this paper.  Indeed,
in the integral of Eq.~\eqref{eq:9} the kernel $K$ gives the same
``weight'' to all points on circles centered in $\vec x$, irrespective
of the specific value of $\rho$ on the various points of the circles
(what matters here is only the \textit{average\/} value of $\rho$ on a
circle, and not the specific value on the various points).  Hence, we
do not expect that this technique can solve the bias introduced by
the correlation between the extinction $A$ and the density of
background stars $\rho$ expressed in Eq.~\eqref{eq:2}, especially if
steep gradients are present in the intrinsic cloud column density.  We
also expect that the bias present in the nearest neighbours
interpolation increases with the number of neighbours $N$.  Indeed,
when $N$ increases, the ``size'' of the kernel $K(r; \vec x)$, i.e. the
range in $r$ where the kernel is significantly different from 0, also
increases because of the effects of the polynomial terms in
Eq.~\eqref{eq:10}.  This, as shown by Eq.~\eqref{eq:9}, implies that
the kernel averages out more distant regions in the sky, where
differences among the local densities can be significant.

In order to better understand the points mentioned in the previous
paragraph, it is useful to look at simple example.  Consider a cloud
characterized by a Heaviside column density:
\begin{equation}
  \label{eq:14}
  A(\vec x) =
  \begin{cases}
    A_1 & \text{if $x_1 < 0 \; ,$} \\
    A_2 & \text{if $x_1 \ge 0 \; .$}
  \end{cases}
\end{equation}
In this case we can basically carry out all the calculations
analytically, and obtain for each value of $x_1$ the expected average
measured extinction $\langle \hat A \rangle$.  We do not report here
the relevant equations, but rather show the results obtained in a
typical case in Fig.~\ref{fig:1}.  As shown by this plot, the nearest
neighbours are clearly biased: for example, the various curves are not
symmetric around $x_1 = 0$, and in addition 
\begin{equation}
  \label{eq:15}
  \langle \hat A \rangle = \frac{A_1 10^{\alpha k_\lambda A_1} + A_2
    10^{\alpha k_\lambda A_2}}{10^{\alpha k_\lambda A_1} + 10^{\alpha
      k_\lambda A_2}} < \frac{A_1 + A_2}{2} \; .
\end{equation}
In order to better understand the bias, we also plot in
Fig.~\ref{fig:1} the expected results in case of a uniform density,
obtained assuming $\alpha = 0$ in Eq.~\eqref{eq:2} (dashed lines): in
this case the curves are perfectly symmetric around $x_1 = 0$ and
$\langle \hat A \rangle = (A_1 + A_2) / 2$.  A comparison of the solid
and dashed lines also confirms that curves with large $N$, being less
steep, are biased on a larger interval around $x_0 = 0$.  This is
expected, since as shown by Eq.~\eqref{eq:10} the intrinsic size of
the smoothing kernel $K$ increases with $N$.

\begin{figure}
  \centering
  \includegraphics[width=\hsize]{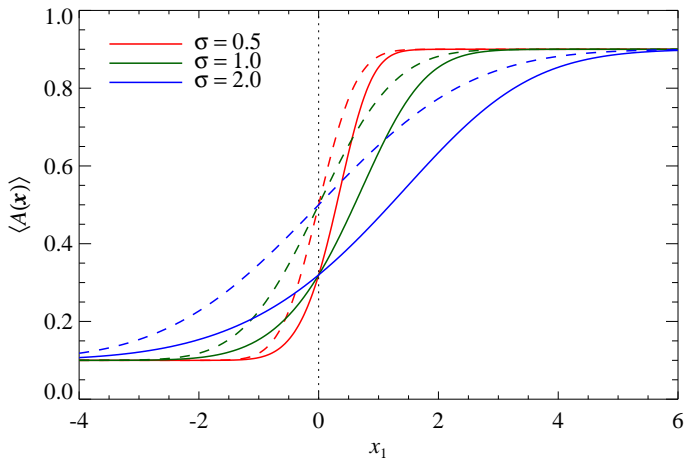}
  \caption{Similarly to Fig.~\ref{fig:1}, but for the moving average
    smoothing with a Gaussian window function with different
    dispersion parameters $\sigma$. As for Figs.~\ref{fig:1} and
    \ref{fig:2}, the solid lines, representing the expected mean
    measurements, are always below the dashed lines, representing the
    same measurements in the ideal case where the density of
    background stars is uniform within the whole field.}
  \label{fig:3}
\end{figure}

One might wanders if the bias decreases by using a median estimator
instead of a simple mean, as done by \citet{2005A&A...435..131C,
  2006A&A...445..999C}.  These authors still identify, for each sky
position $\vec x$, the $N$ nearest neighbour stars, and then evaluate
a simple median of the relative extinction measurements.  Luckily, it
is possible to evaluate the exact statistical properties of the median
estimator (see \citealp{2002A&A...395..733L}, Appendix~A).  For this
purpose, it is useful to evaluate the probability distribution $p_N(A;
\vec x)$ that one of the $N$ neighbours around the position $\vec x$
has a column density $A$:
\begin{equation}
  \label{eq:16}
  p_N(A; \vec x) = \int \diff \mu \, \mathrm{e}^{-\mu}
  \frac{\mu^{N-2}}{N!} \int_{B(\mu; \vec x)} \diff x' \, \rho(\vec x')
  \delta \bigl( A - A(\vec x) \bigr) \; ,
\end{equation}
where $B(\mu; \vec x)$ is the disk centered at $\vec x$ and with
radius $r$ defined by $\mu = \mu(r; \vec x)$ (note that since $\mu(r;
\vec x)$, for $\vec x$ fixed, is a monotonic function of $\vec x'$
this definition is unique).  The cumulative distribution associated
with $p_N(A; \vec x)$ is given by
\begin{equation}
  \label{eq:17}
  P_N(A; \vec x) = \int \diff \mu \, \mathrm{e}^{-\mu}
  \frac{\mu^{N-2}}{N!} \nu(\mu, A; \vec x) \; .
\end{equation}
In this equation, $\nu(\mu, A; \vec x)$ is the integral of $\rho(\vec
x)$ carried out over all points of $B(\mu; \vec x)$ where $A(\vec x')
< A$:
\begin{equation}
  \label{eq:18}
  \nu(\mu, A; \vec x) \equiv \int_{B(\mu; \vec x) \cap \{ \vec x' |
    A(\vec x') < A \}} \rho(\vec x') \, \diff x' \; .
\end{equation}
Note that $\nu(\mu, A; \vec x) \rightarrow \mu$ as $A \rightarrow
\infty$.  Finally, the median of the $N = 2k - 1$ nearest neighbours
is then provided by the expression
\begin{equation}
  \label{eq:19}
  p_\mathrm{m}(A) = k {N \choose k} p_N(A; \vec x) \bigl[P_N(A; \vec
  x) \bigr]^{k-1} \bigl[ 1 - P_N(A; \vec x) \bigr]^{k-1} \; .
\end{equation}

Although it is difficult to obtain a general expression for the bias
introduced by the median-nearest neighbours estimator, it is clear
that to first approximation the same arguments discussed above for the
simple mean apply (note also that, trivially, when $N = 1$ the two
estimators are identical).  However, for the simple example considered
above, we can actually carry out the calculations using
Eqs.~(\ref{eq:16}--\ref{eq:19}) and show that the bias is still
significant.  Figure~\ref{fig:2} summarizes the results obtained in
the toy model described by Eq.~\eqref{eq:14}, and proves that no real
improvement is obtained from the use of a median estimator (cf.\
Fig.~\ref{fig:1}).  In particular, a severe bias is still present, and
its amount increases with $N$.  For comparison, in Fig.~\ref{fig:3} we
also show the results obtained from the simple moving average
smoothing discussed in Sect.~\ref{sec:moving-aver-smooth}:
interestingly, the plot is very similar to the one for a the simple
average nearest neighbors interpolation (Fig.~\ref{fig:1}).  This, in
reality, is not too much surprising because the expected average
values $\bigl\langle A(x) \bigr\rangle$ measured with these two
techniques can be described in terms of a simple convolution with some
appropriate kernels [cf. Eq.~\eqref{eq:4} and \eqref{eq:10}; see also
\citealp{2002A&A...395..733L}].

In summary, a careful analysis of the nearest neighbors shows that
this interpolation technique, for the specific case of extinction
measurements, is still affected by a significant bias.  At least for
the case of a simple mean estimator, the origin of the problem can be
traced to the different behaviour of the kernel $K(\vec x; \vec x')$
with respect to the density $\rho(\vec x')$ in Eq.~\eqref{eq:9}: in
particular, the kernel $K$ does not respond directly to variations of
$\rho$, and instead depends only $\mu(\vec x'; \vec x)$, i.e.\ on the
total ``mass'' within a disk of radius $r = \langle \vec x - \vec x'
\rangle$.

\section{\textsc{Nicest}: over-weighting high column-density
  measurements}
\label{sec:over-weighting-high}

As shown by Eq.~\eqref{eq:8}, the bias discussed here strongly depends
on the small-scale structure of the molecular cloud through the
probability distribution $p_A(A)$.  Although on large scales many
models predict a log-normal distribution for $p_A$, clearly we should
expect significant deviations from the theoretical expectations on the
small scales often investigated in NIR studies.  In addition, even at
large scales the superposition of different cloud complexes, or a
significant ``thickness'' of a cloud along the line of sight, can
produce significant deviations from a log-normal distribution
(\citealp{2001ApJ...557..727V}; see also \citealp{2006A&A...454..781L}
for a case where the log-normal distribution is not a good
approximation). Hence, \textit{we need to be able to correct for the
  substructure bias in a model independent way}.

\subsection{Statistical properties of the moving average smoothing}
\label{sec:stat-prop-moving}

In order to better understand, and then fix, the bias present in
Eq.~\eqref{eq:1} in the case of the moving average smoothing, we use
the theory developed in \citet{2001A&A...373..359L,
  2002A&A...392.1153L, 2003A&A...407..385L}.  The key equations to
obtain the ensemble average of $\hat A(\vec x)$ are summarized below:
\begin{align}
  \label{eq:20}
  Q(s; \vec x) = {} & \int \bigl[ \mathrm{e}^{-s w(\vec x'; \vec x)} -
  1 \bigr] \rho(\vec x') \, \diff x' \\
  \label{eq:21}
  Y(s; \vec x) = {} & \exp \bigl[ Q(s; \vec x) \bigr] \; , \\
  \label{eq:22}
  C(w; \vec x) = {} & \frac{1}{1 - P(x)} \int_0^\infty \mathrm{e}^{-w
    s} Y(s; \vec x) \, \diff s \; , \\
  \label{eq:23}
  \bigl\langle \hat A(\vec x) \bigr\rangle = {} & \int A(\vec x')
  w(\vec x'; \vec x) C\bigl( w(\vec x'; \vec x), \vec x \bigr)
  \rho(\vec x') \, \diff x' \; .
\end{align}
Let us briefly comment this set of equations:
\begin{itemize}
\item The equations are invariant upon the transformation $w(\vec x';
  \vec x) \mapsto q w(\vec x'; \vec x)$, with $q$ positive constant.
  This is expected, since an overall multiplicative constant in the
  weight function does not effect Eq.~\eqref{eq:1}.  For simplicity,
  in the following we will assume, without loss of generality, that
  the weight function is normalized according to
  \begin{equation}
    \label{eq:24}
    \int w(\vec x'; \vec x) \rho(\vec x') \, \diff x' = 1 \; .
  \end{equation}
\item The quantity $Q(s; \vec x)$ can be shown to be related to the
  Laplace transform of $p_w(w; \vec x)$, the probability distribution
  for the values of the weight function $w(\vec x'; \vec x)$ with
  $\vec x$ fixed.
\item $P(\vec x)$ is the probability that no single star is present
  within the support $\pi_w(\vec x)$ of $w(\vec x'; \vec x)$, defined
  as $\pi_w(\vec x) \equiv \{ x' \,|\, w(\vec x'; \vec x) > 0 \}$.  Hence,
  $P(\vec x)$ can be simply evaluated from
  \begin{equation}
    \label{eq:25}
    P(\vec x) = \exp \left[ - \int_{\pi_w(\vec x)} \rho(\vec x') \diff
    x' \right] \; .
  \end{equation}
  Note that $P(\vec x) = 0$ for weight functions with infinite
  support (such as a Gaussian). 
\item $C(w; \vec x)$ is a simple Laplace transform of $Y(s; \vec x)$.
  This is the key quantity that enters the final result \eqref{eq:23}
  together with the original smoothing function $w(\vec x'; \vec x)$
  and the star density $\rho(\vec x')$.
\item A key parameter in the calculation of $C(w)$, and thus on the
  final result $\langle \hat A(\vec x) \rangle$, is the so-called
  \textit{weight number of objects\/} $\mathcal{N}(\vec x)$, defined
  as
  \begin{equation}
    \label{eq:26}
    \mathcal{N}(\vec x) \equiv \left[ \bigl[ 1 - P(\vec x) \bigr] \int
    \bigl[ w(\vec x'; \vec x) \bigr]^2 \rho(\vec x') \, \diff x'
  \right]^{-1} \; ,
  \end{equation}
  where as usual $w(\vec x'; \vec x)$ is taken to be normalized
  according to Eq.~\eqref{eq:24}.  Informally, $\mathcal{N}(\vec x)$
  counts the number of stars that contribute (with a weight
  significantly different from zero) to the average $\hat A(\vec x)$.
\item In the limit $\mathcal{N}(\vec x) \gg 1$ it is possible to
  obtain a simple expression for $C(w; \vec x)$, which for $P(\vec x)
  = 0$ takes the form
  \begin{equation}
    \label{eq:27}
    C(w; \vec x) \simeq \frac{1}{1 + w} + \frac{\bigl[
      \mathcal{N}(\vec x) \bigr]^{-1}}{(1 + w)^3} \; .
  \end{equation}
  This expression shows that to lowest order $C \simeq 1 - w +
  \mathcal{N}^{-1}$, where both terms $w$ and $\mathcal{N}^{-1}$
  introduce a \textit{relative\/} correction of order
  $\mathcal{N}^{-1}$ in the final result $\bigl\langle \hat A(\vec x)
  \bigr\rangle$.
\end{itemize}

The results summarized above rigorously prove the statements of
Sect.~\ref{sec:moving-aver-smooth}.  In particular, in the limit of a
large weight number of objects $\mathcal{N}(\vec x)$, we have $C(w)
\simeq 1$, and thus Eq.~\eqref{eq:23} together with the normalization
\eqref{eq:24} reduces to Eq.~\eqref{eq:4}.  In practice, numerical
simulations show that Eq.~\eqref{eq:4} is already accurate for
relatively small weight numbers (of the order of $\mathcal{N} \sim
10$; see Fig.~3 of \citealp{2001A&A...373..359L}).

\subsection{A fix for the inhomogeneities bias}
\label{sec:fix-inhom-bias}

The bias of Eq.~\eqref{eq:1} is essentially due to a change in the
density of extinction measurements as a function of $A$
[Eq.~\eqref{eq:2}].  In the framework of the moving average smoothing,
it is thus reasonable to try to ``compensate'' this effect by
including in the weights of the estimator \eqref{eq:1} a factor
$10^{\alpha k_\lambda A}$:
\begin{equation}
  \label{eq:28}
  w_n = w(\vec x_n; \vec x) 10^{\alpha k_\lambda \hat A_n} \; .
\end{equation}
With this modification, we expect $\hat A$ to be unbiased provided
that $\mathcal{N}$ is large (so that we can take $C \simeq 1$) and
that measurement errors on $\hat A_n$ can be neglected.  Instead, in
presence of significant errors on the individual column density
estimations, we expected $\hat A$ to be biased toward large
extinctions.  This happens because of the non-linearity of the
introduced factor in $\hat A_n$: for example, $\hat A_n$ is
symmetrically distributed around $A_n$, $10^{\alpha k_\lambda \hat
  A_n} \hat A_n$ will be skewed against values larger than $10^{\alpha
  k_\lambda A_n} A_n$.  In summary, the modified weight of
Eq.~\eqref{eq:28} removes (to a large degree) the bias due to the
cloud substructure, but introduces a new bias related to the scatter
of the individual extinction measurements.  While the former is
basically unpredictable, the latter can be estimated accurately
provided we know the expected errors on $\hat A_n$.

\begin{figure}
  \centering
  \includegraphics[width=\hsize]{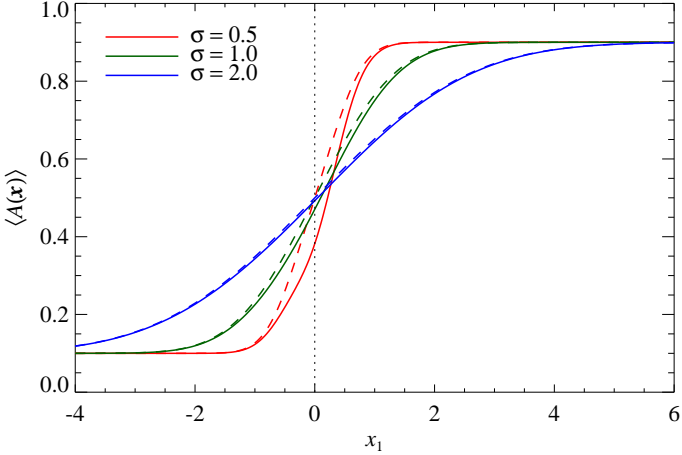}
  \caption{Similarly to Fig.~\ref{fig:3}, but for the \textsc{Nicest}
    moving average smoothing (see
    Sect.~\ref{sec:over-weighting-high}).  In this case a significant
    bias is observed only for $\sigma = 0.5$, which however correspond
    to a very small weight number $\mathcal{N} \simeq 2$.}
  \label{fig:4}
\end{figure}

In order to better understand the properties of the estimator
\eqref{eq:1} with the new weight, let us evaluate the average
$\bigl\langle \hat A \bigr\rangle$.  For simplicity, initially we will
ignore measurement errors and focus on the effects introduced by a
non-uniform density.  We first note that Eq.~\eqref{eq:28} is
equivalent to the use of a weight $w_n = w'(\vec x_n; \vec x)$, where
\begin{equation}
  \label{eq:29}
  w'(\vec x'; \vec x) \equiv w(\vec x'; \vec x) / \rho(\vec x')
  \; .
\end{equation}
We require the new weight $w'(\vec x'; \vec x)$ to be normalized
according to Eq.~\eqref{eq:24}, which implies
\begin{equation}
  \label{eq:30}
  \int w(\vec x'; \vec x) \, \diff x' = 1 \; .
\end{equation}
Interestingly, the normalization condition \eqref{eq:30} for $w$ does
not depend on the density $\rho$ anymore, and it is thus possible to
choose for $w$ a spatially invariant function $w(\vec x'; \vec x) =
w(\vec x' - \vec x)$ (a typical choice would be a Gaussian, $w \propto
\mathrm{e}^{-|\vec x' - \vec x| / 2 \sigma^2}$).  We have then
\begin{align}
  \label{eq:31}
  \bigl\langle \hat A(\vec x) \bigr\rangle = {} & \int A(\vec x')
  w'(\vec x'; \vec x) \bigl[ 1 - w'(\vec x'; \vec x) +
  \mathcal{N}^{-1}(\vec x) \bigr] \rho(\vec x') \, \diff x' \notag\\
  {} = {} & \int A(\vec x') w(\vec x'; \vec x) \bigl[ 1 - w(\vec x';
  \vec x) / \rho(\vec x') + \mathcal{N}^{-1}(\vec x) \bigr] \, \diff x' \; .
\end{align}
After a few manipulations we can rewrite this expression in the form
\begin{align}
  \label{eq:32}
  \bigl\langle \hat A(\vec x) \bigr\rangle = {} & \bar A(\vec x) -
  \frac{1}{\rho_0} \int w^2(\vec x'; \vec x) 10^{\alpha k_\lambda
    A(\vec x')} \bigl[ A(\vec x') - \bar A(\vec x) \bigr]
  \, \diff x' \notag\\
  {} \simeq {} & \bar A(\vec x) - \frac{10^{\alpha k_\lambda \bar
      A(\vec x)}}{\rho_0} \int w^2(\vec x'; \vec x) \bigl[
  \Delta(\vec x'; \vec x) \notag\\
  & \phantom{\bar A(\vec x) - \frac{10^{\alpha k_\lambda \bar A(\vec
        x)}}{\rho_0} \int w^2(\vec x'; \vec x) \bigl[} + \beta
  \Delta^2(\vec x'; \vec x) \bigr] \, \diff x \; ,
\end{align}
where in the second equality we have expanded the exponential term to
first order using the definitions following Eq.~\eqref{eq:8}.  
In summary, the bias $\bigl\langle \hat A(\vec x) \bigr\rangle - \bar
A(\vec x)$ of the proposed estimator is composed of two terms:
\begin{itemize}
\item A weighted average of $\Delta(\vec x'; \vec x)$.  Since the
  weighted average uses $w^2(\vec x'; \vec x)$ as weight, this term is
  not expected to vanish identically unless the weight is a top-hat
  function [cf.\ Eq.~\eqref{eq:8}, where instead the weighted average
  involves $w(\vec x'; \vec x)$ and therefore no linear contribution
  in $\Delta$ is present].  Still, we do not expect any systematic
  bias related to this term, because on average $\Delta(\vec x'; \vec
  x)$ by construction has alternating sign.
\item A weighted average of $\Delta^2(\vec x'; \vec x)$.  This term is
  manifestly negative and therefore introduces a bias in the result.
  The bias is very similar to the original bias discussed in
  Eq.~\eqref{eq:8}: it is proportional to $\beta \equiv \alpha
  k_\lambda \ln 10$, and depends on the scatter $\Delta(\vec x'; \vec
  x) \equiv A(\vec x') - \bar A(\vec x)$ of the local column density
  with respect to its average value.  However, there is a significant
  difference: the bias is inversely proportional to $\rho_0
  10^{-\alpha k_\lambda \bar A(\vec x)}$, i.e. to the local average
  density within the weight function; moreover, the average is taken
  over $w^2$ instead of $w$.  These two differences, together,
  indicate that the bias of Eq.~\eqref{eq:32}, compared to the on of
  Eq.~\eqref{eq:8}, is smaller by a factor $\sim \mathcal{N}(\vec x)$.
\end{itemize}
In other words, the modification of the weight operated by
Eq.~\eqref{eq:28} reduces the bias present in the original simple
moving weight average by a factor $\mathcal{N}(\vec x)$, which in
typical cases is significantly larger than unity.\footnote{In reality,
  a detailed calculation shows that the key parameter here is
  $\mathcal{N}_\mathrm{eff}(\vec x)$, which is defined analogously to
  $\vec{N}(\vec x)$ but with $w$ replaced by $w_\mathrm{eff} \equiv w
  C(w)$.  As shown in \citet{2002A&A...395..733L},
  $\mathcal{N}_\mathrm{eff} \ge 1$, so the bias never increases.}\@ On
the other hand, the fact that the method is ineffective then
$\mathcal{N}$ is low is evident from the definition of $w_n$: in
particular, as $\rho \rightarrow 0$ we expect that only a single star
contributes to the averages of Eq.~\eqref{eq:1} with a weight $w_n =
w(\vec x_n; x) 10^{\alpha k_\lambda \hat A_n}$ significantly different
from zero, but then trivially $\hat A = \hat A_n$ and no correction is
effectively performed.\footnote{In this example $\mathcal{N}
  \rightarrow 0$ while $\mathcal{N}_\mathrm{eff} \rightarrow 1$, which
  agrees with statement of the previous footnote.}

In Fig.~\ref{fig:4} we show the effects of the proposed smoothing
technique on the toy-model discussed in
Sect.~\ref{sec:nearest-neighbors}.  A comparison with Fig.~\ref{fig:3}
shows that the \textsc{Nicer} interpolation is very effective in
reducing the bias of the simple moving average interpolation,
especially for relatively large values of the smoothing factor
$\sigma$ (and thus of the weight number of objects $\mathcal{N}$).

We now consider the effects of measurement errors in the modified
estimator.  For simplicity, we consider the limit of a large number of
stars ($\mathcal{N} \gg 1$), so that $C(w) \simeq 1$ (in any case, as
shown above, the technique proposed is inefficient when $\mathcal{N}$
is of order of unity).  In this case, a simple calculation shows that
measurements errors introduce a bias $B_\mathrm{err}$ of the order
\begin{equation}
  \label{eq:33}
  B_\mathrm{err} \simeq \beta \frac{\sum_n w_n \sigma_n^2}{\sum_n w_n}
  \; , 
\end{equation}
where $\{ \sigma^2_n \}$ are the variances on $\{ \hat A_n \}$.  In
other words, the method proposed in this paper introduces a small bias
toward large extinction values.  In order to estimate the order of
magnitude of $B_\mathrm{err}$, we note that the median error in $K$
band extinctions on \textsc{Nicer} column density estimates from the
2MASS catalog is typically\footnote{The expected error on $\hat A_n$
  is calculated using the standard \textsc{Nicer} techniques, and thus
  includes both the typical 2MASS photometric uncertainties and the
  star color scatters as measured in control fields where the
  extinction is negligible.  Hence, this error includes the effects of
  different stellar populations on the scatter in the intrinsic star
  colors, as long as the various stellar populations are represented
  in the control field.} $\sigma \simeq 0.13 \mbox{ mag}$, and that
less than $1\%$ of stars have $\sigma > 0.2 \mbox{ mag}$.  Taking
$\sigma_n = 0.13 \mbox{ mag}$ for all stars in Eq.~\eqref{eq:33}, we
find a bias $\langle \hat A \rangle - A \simeq 0.02 \mbox{ mag}$
(always in the $K$ band).  Interestingly, although very small this
bias, in contrast to the one described by Eq.~\eqref{eq:32}, can be
corrected to first order because its expression involves only known
quantities.

In summary, we propose to replace the estimator \eqref{eq:1} with
\begin{equation}
  \label{eq:34}
    \boxed{\hat A = \frac{\sum_n w_n\hat
        A_n}{\sum_n w_n } - 
      \beta \, \frac{\sum_n w_n \sigma^2_n}{\sum_n w_n} \; ,}
\end{equation}
where $w_n \equiv w(\vec x'; \vec x) 10^{\alpha k_\lambda \hat A_n}$.
This new estimator, that we name \textsc{Nicest}, significantly
reduces the bias due small-scale inhomogeneities in the extinction map
and is (to first order) unbiased with respect to uncertainties on
$\hat A_n$.

If one is ready to accept a small bias on $\hat A$ for extinction
measurement uncertainties, it is also possible to use
Eq.~\eqref{eq:34} without the last term.  In this respect, we also
note that the correcting factor present in Eq.~\eqref{eq:34} is
independent of the local extinction, and only depends on the
individual errors on the measurements (these typically are a
combination of the photometric uncertainties and of the intrinsic
scatter of colors of the stellar population considered).  As a result,
it is not unrealistic to expect that this factor is approximately
constant within the field of observation, and that it is equally
present in the control field used to fix the zero of the extinction.

Finally, we note that in \textsc{Nicest} a central role is played by
the slope of the number counts $\alpha$, operationally defined through
Eq.~\eqref{eq:2}.  In reality, however, it clear that Eq.~\eqref{eq:2}
cannot strictly hold, because there is an upper limit on the
(un-reddened) magnitude of stars.  Hence, depending on the limiting
magnitude of the observations, there is an upper limit to the
measurable extinction of stars.  As a result, in regions where the
extinction is larger than this limit, the density of background stars
vanishes, with the result that in these cases even \textsc{Nicest}
cannot be unbiased.  The exact value of the upper measurable
extinction depends on many factors, including the waveband $\lambda$
and limiting magnitude $m_\mathrm{max}$ of the observations, and the
distance $d$ of the molecular cloud.  An approximate relation for the
maximum extinction $A_\mathrm{max}$ is
\begin{equation}
  \label{eq:35}
  A_\mathrm{max} \sim \frac{1}{k_\lambda} \left[ m_\mathrm{max} - 
    M_\mathrm{max} - 5 \log_{10} \left( \frac{d}{10 \mbox{ pc}} \right)
  \right] \; ,
\end{equation}
where $M_\mathrm{max}$ is the maximum absolute magnitude of stars in
the band considered.  The maximum extinction $A_\mathrm{max}$ should
be evaluated in each specific case; for example, for typical 2MASS
observations, using $m_\mathrm{max} \simeq 15 \mbox{ mag}$, $k_H =
1.55$ \citep{2005ApJ...619..931I}, and $M_\mathrm{max} \simeq -5
\mbox{ mag}$ we obtain $A_\mathrm{max} \simeq 10 \mbox{ mag}$ for a
cloud located at $100 \mbox{ pc}$.

\subsection{Foreground stars}
\label{sec:foreground-stars}

\begin{figure*}
  \centering
  \includegraphics[width=0.49\hsize, bb=149 305 485 524]{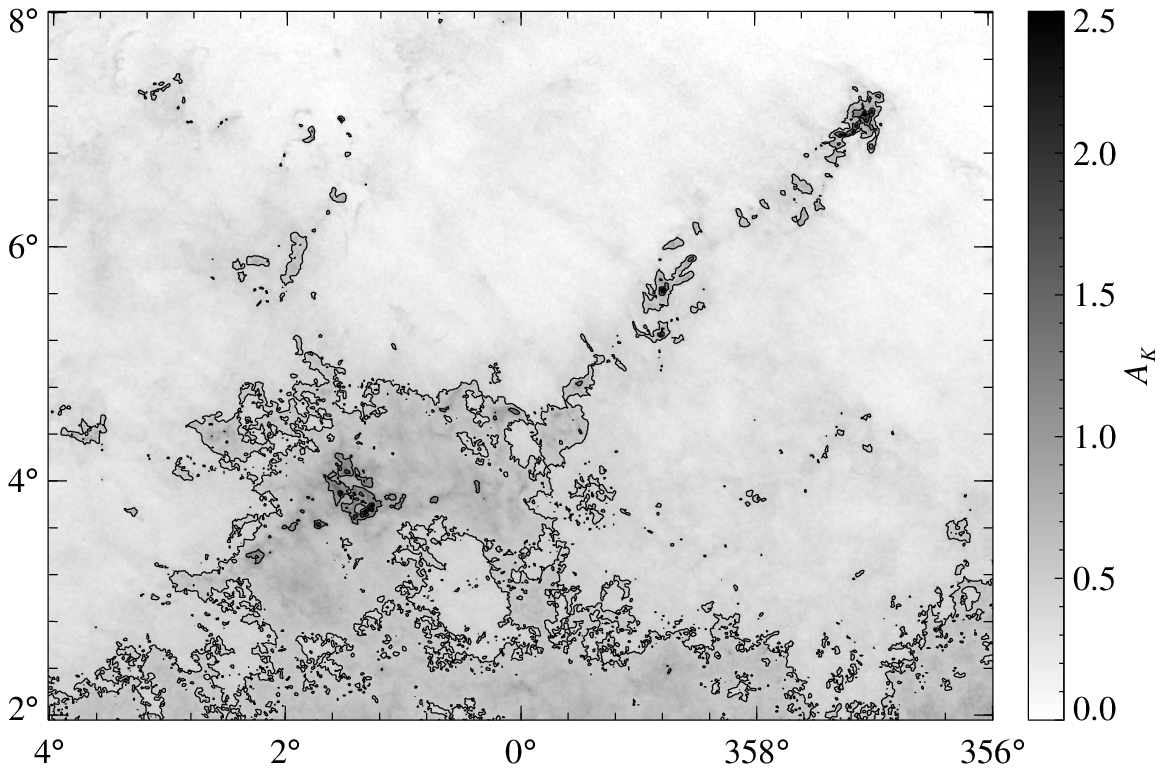}
  \hfill
  \includegraphics[width=0.49\hsize, bb=149 305 485 524]{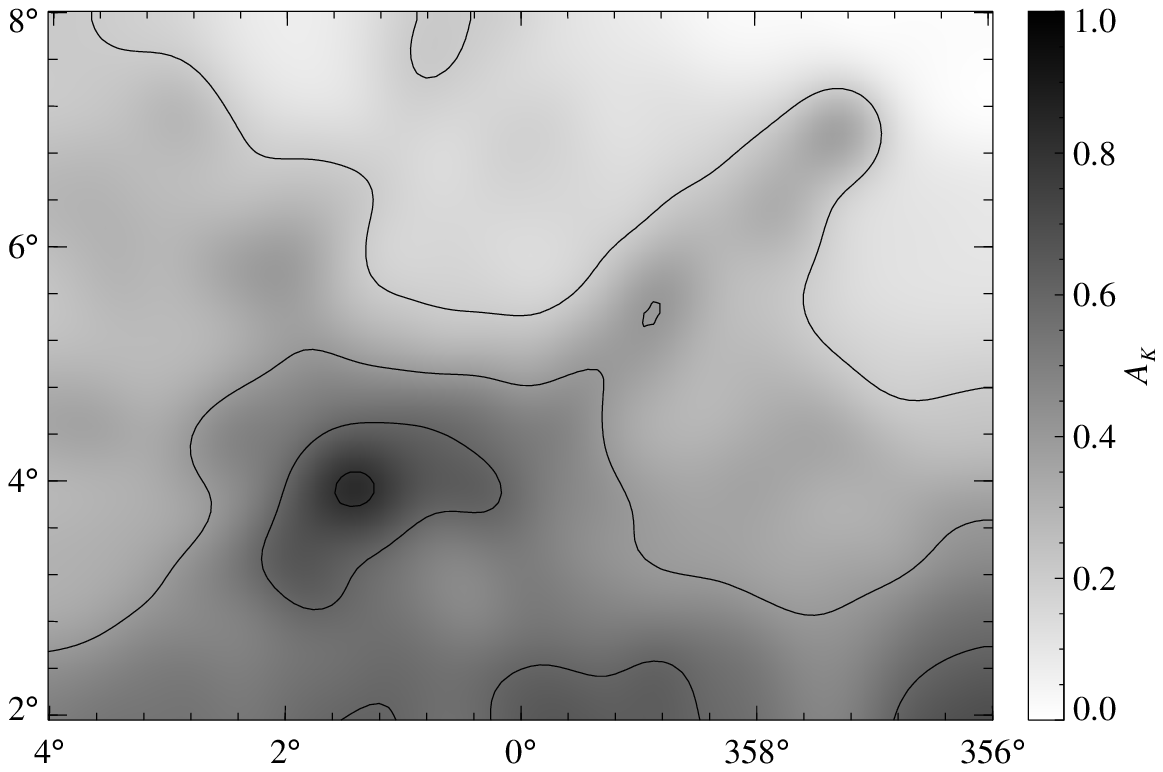}
  \vskip 1em
  \includegraphics[width=0.49\hsize, bb=149 305 485 524]{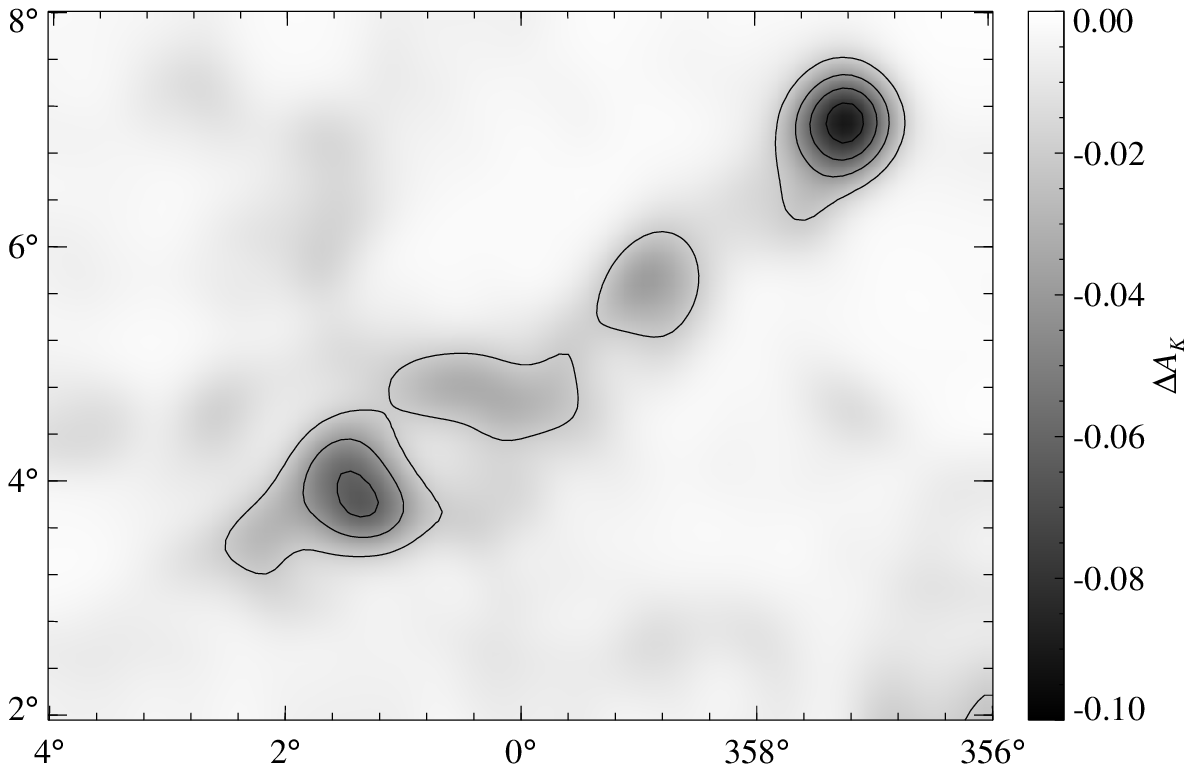}
  \hfill
  \includegraphics[width=0.49\hsize, bb=149 305 485 524]{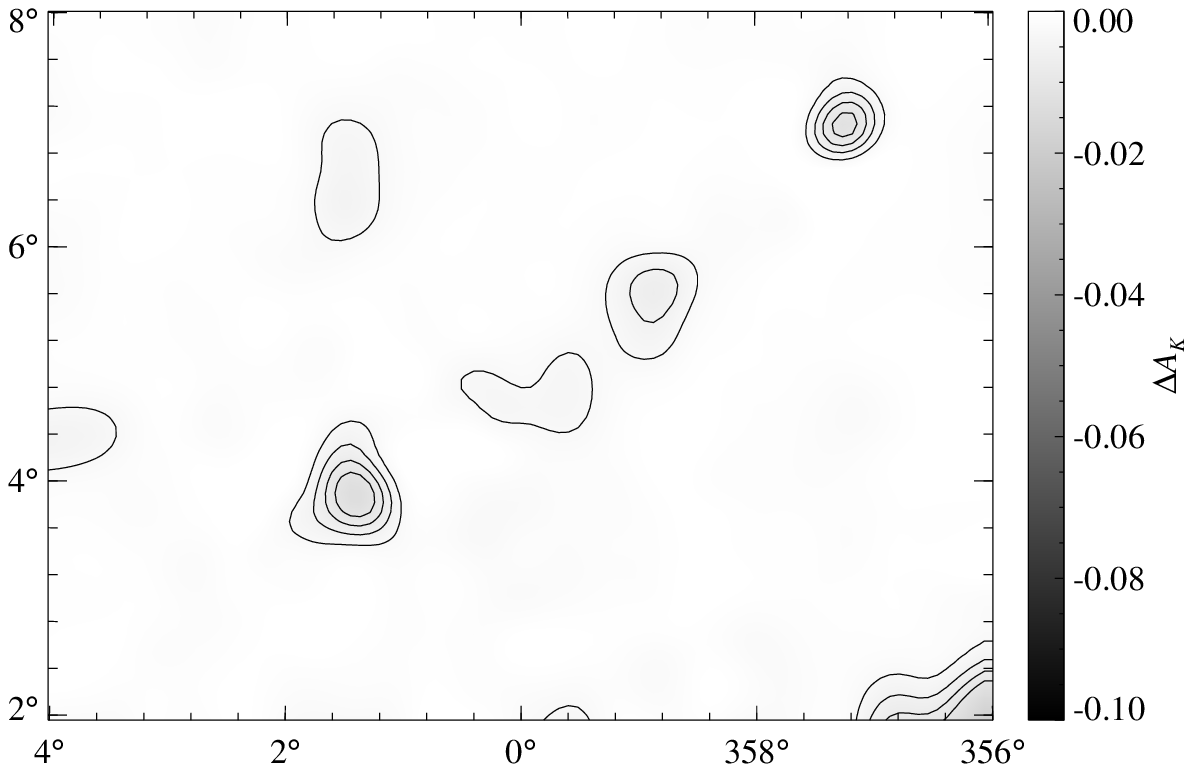}
  \caption{\textbf{Top left.} The original 2MASS/\textsc{Nicer} map of
    the Pipe nebula used in the simulations. \textbf{Top right.} The
    mean reconstructed map that one would obtain if the density of
    background star were uniform.  This is basically a smoothed
    version of the original map in the left. Levels are spaced at $0.2
    \mbox{ mag}$.  \textbf{Bottom left.} The difference between the
    average reconstructed maps, calculated from background stars
    following the density of Eq.~\eqref{eq:2}, and the average map at
    top right.  Levels are spaced at $0.02 \mbox{ mag}$.
    \textbf{Bottom right.} Same map as bottom left, but using the
    improved method presented in this paper.  Levels are spaced at
    $0.0025 \mbox{ mag}$.}
  \label{fig:5}
\end{figure*}

Foreground stars do not contribute to the extinction signal, but do
contribute to the noise of the estimators \eqref{eq:1} and
\eqref{eq:34}, and thus whenever possible they should be excluded from
the analysis.  Usually, foreground stars are easily identified in high
($A_K > 1 \mbox{ mag}$) column density regions, but are almost
impossible to distinguish in low and mid extinction regions.  So far
we assumed that all stars are background to the cloud, but clearly in
real observations we should expect a fraction $f$ of foreground
objects.  In principle, if this fraction is known, we could correct
the column density estimate $\hat A$ into $\hat A / (1 - f)$: in this
way, we would obtain an unbiased estimator of the column density at
the price of an increased noise (due to foreground objects).  In
practice, the fraction of foreground stars is itself an increasing
function of the local column density, i.e.\ $f = f(A)$, because of the
decreased density of background stars in highly extincted regions.
Hence, the simple scheme proposed above is not easily implemented.

\begin{figure*}
  \centering
  \includegraphics[width=0.49\hsize, bb=149 305 485 524]{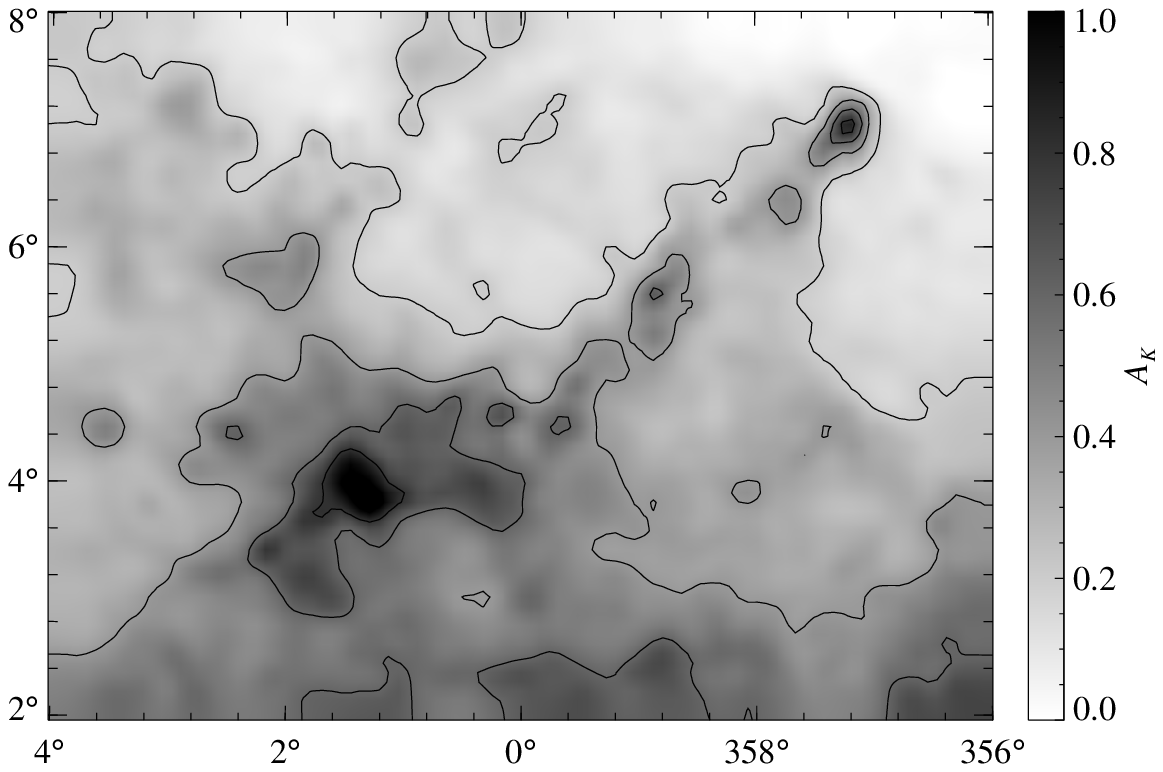}
  \hfill
  \includegraphics[width=0.49\hsize, bb=149 305 485 524]{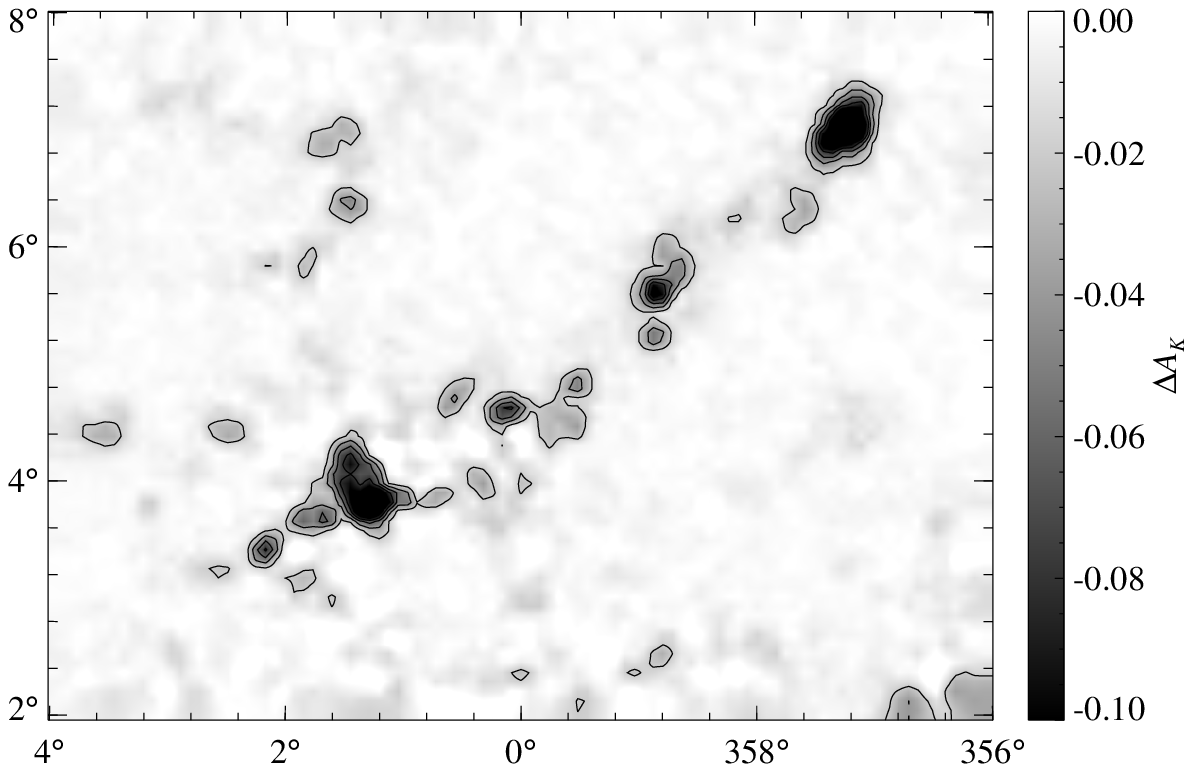}
  \caption{\textbf{Left.} Average reconstructed map using a 1-star ($N
    = 1$) Voronoi method, assuming a uniform distribution of stars.
    Levels are spaced at $0.2 \mbox{ mag}$. \textbf{Right.} Difference
    between the average reconstructed Voronoi map, obtained from
    background stars following the density of Eq.~\eqref{eq:2}, and
    the average true map to the left. Levels are spaced at $0.02
    \mbox{ mag}$}
  \label{fig:6}
\end{figure*}

Interestingly, \textsc{Nicest} perfectly adapts to the presence of
foreground stars.  Suppose that in regions \textit{with negligible
  extinction\/} a fraction $f_0 \equiv f(A = 0) < 1$ of stars is
foreground to the cloud: then, in our notation, we can rephrase this
by assigning a non-vanishing probability to $p_A(A = 0)$, i.e.\ by
treating foreground stars as a special case of substructure (much like
``holes'' in the cloud).  As we consider higher density regions, the
fraction $f(A)$ of foreground stars increases but, because of the
correcting factor in $w_n$, we still expect to measure $(1 - f_0) A$:
in other words, the estimator $\hat A / (1 - f_0)$, with $\hat A$
given by Eq.~\eqref{eq:34} is expected to be unbiased both for
small-scale inhomogeneities and for foreground contamination.  Note
that correcting factor $(1 - f_0)^{-1}$ is usually very close to $1$
for nearby molecular clouds, and can be easily evaluated by comparing
the density of foreground stars (easily measured in highly extincted
regions) with the total density of stars in regions free from
extinction.

\section{Simulations}
\label{sec:simulations}

\begin{figure*}
  \centering
  \includegraphics[width=0.49\hsize, bb=149 305 485 524]{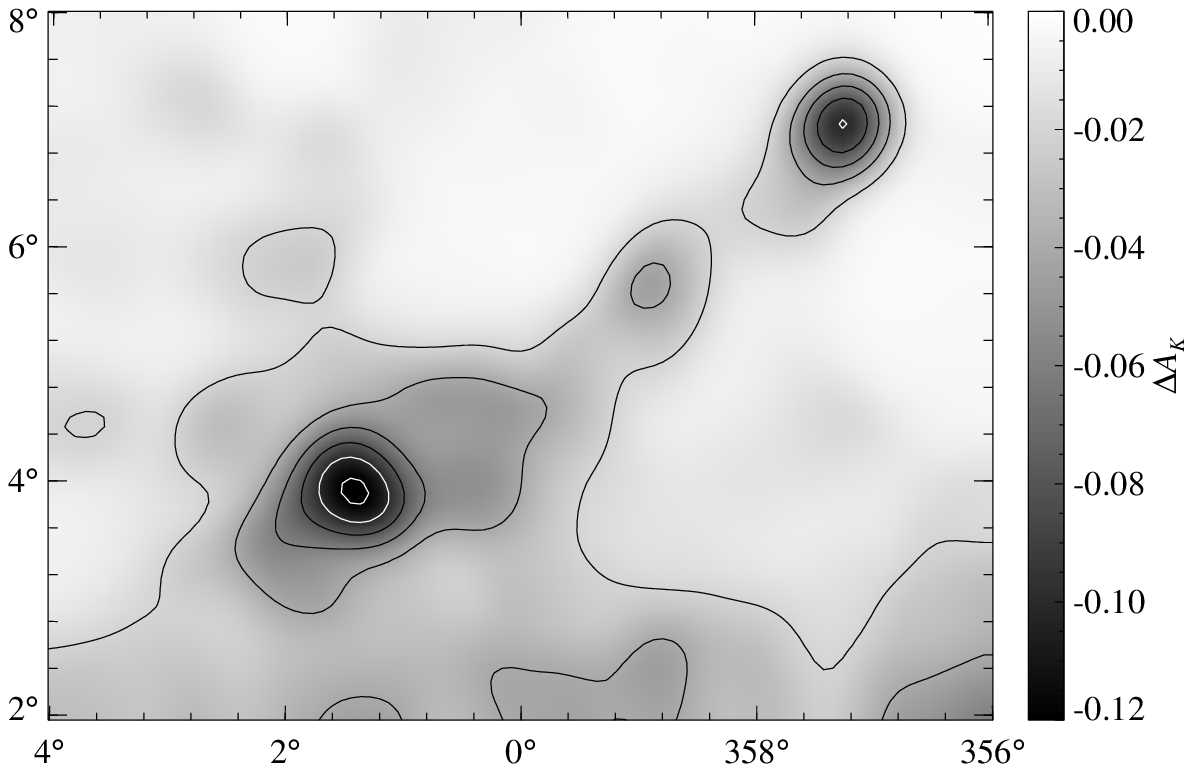}
  \hfill
  \includegraphics[width=0.49\hsize, bb=149 305 485 524]{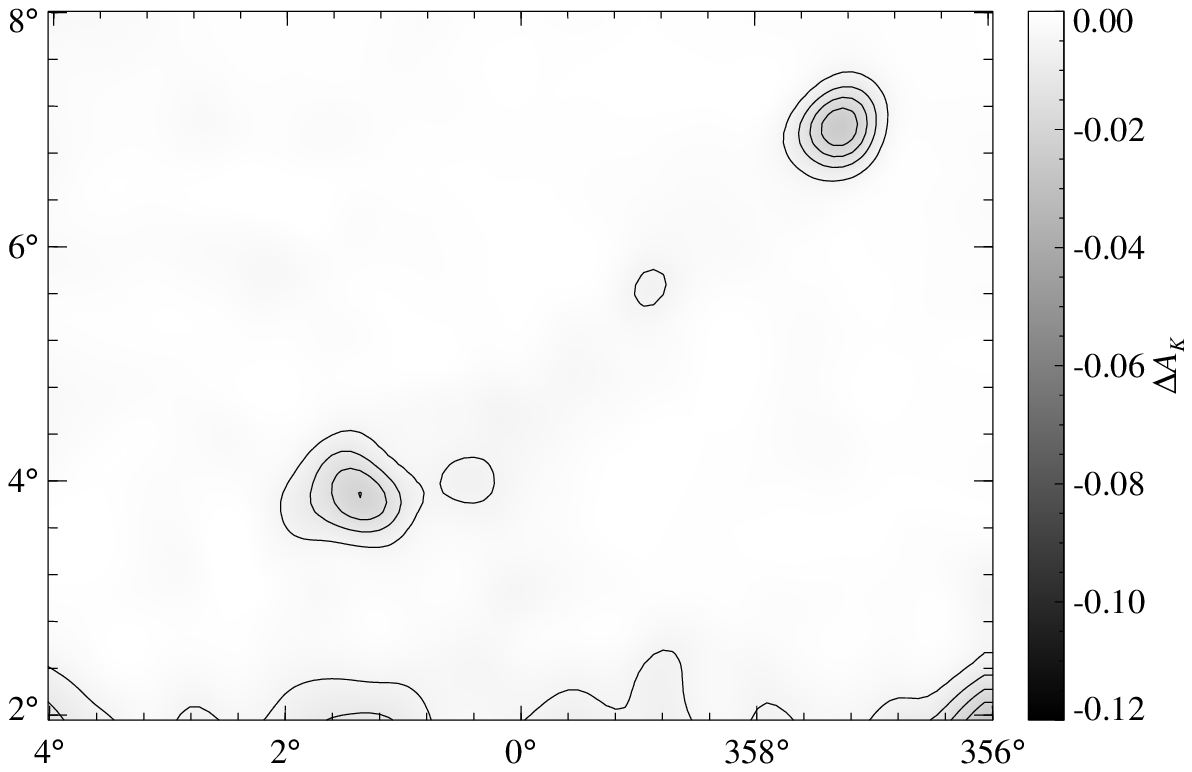}
  \caption{Same as Fig.~\ref{fig:5} (bottom), but with
    a fraction of $f = 0.05$ of foreground stars.  The increases of
    relative number of foreground stars in the denser regions of the
    cloud makes the bias more pronounced.  Still, \textsc{Nicest}
    performs very well.  Contours levels are spaced at $0.02 \mbox{
      mag}$ in the \textsc{Nicer} plot (left) and at $0.005
    \mbox{ mag}$ in the \textsc{Nicest} plot (right); contours at
    $-0.1 \mbox{ mag}$ or below are plotted in white.}
  \label{fig:7}
\end{figure*}

In order to test the effects of substructures and the reliability of
the method proposed here in a real case scenario we performed a series
of simulations.  We started from a 2MASS/\textsc{Nicer} map of the
Pipe nebula \citep{2006A&A...454..781L}, a nearby complex of molecular
clouds seen in projection to the Galactic bulge, an ideal case for
extinction studies \citep[see also][]{BlueBook, 1999PASJ...51..871O}.
We took this map as the \textit{true\/} dust column density of a cloud
complex, and simulated a set of background stars.  In order to show
the effects of substructures, we used a very small density of
background stars ($25 \mbox{ stars deg}^{-2}$ instead of the original
$\sim 9.4 \times 10^{5} \mbox{ stars deg}^{-2}$ used to build the
2MASS/\textsc{Nicer} map).  This, effectively, correspond to a linear
downsize of the structures of the Pipe nebula by a factor $\sim 60$.

The simulations were carried out using the following simple technique.
We generated a set of background stars uniformly distributed on the
field of view.  The stars were characterized by exponential number
counts with exponent $\alpha = 0.34$ in the $K$ band and by intrinsic
color and color scatters similar to the ones observed in the 2MASS
catalog (see Table~2 of \citealp{2005A&A...438..169L} for a complete list of
the parameters used).  We added to each star magnitude the local value
of the of the 2MASS/\textsc{Nicer} extinction and also some random
measurement errors:
\begin{align}
  \label{eq:36}
  \hat J_n = {} & J_n + k_K A_K(\vec x_n) +
  \varepsilon^{(J)}_n \; , \\
  \label{eq:37}
  \hat H_n = {} & H_n + k_H A_K(\vec x_n) +
  \varepsilon^{(H)}_n \; , \\
  \label{eq:38}
  \hat K_n = {} & K + A_k(\vec x_n) + \varepsilon^{(K)}_n \; ,
\end{align}
where $\varepsilon^{(J,H,K)}_n$ denote random variables used to model
the photometric errors of the stars, and where as usual we used the
hat to denote measured quantities.  For simplicity, here, we took the
errors as normal distributed random variables with constant variance
$0.05 \mbox{ mag}$ in all bands (the typical median variance of 2MASS
magnitudes); moreover, we assumed a hard completeness limit at $15
\mbox{ mag}$ in all bands (i.e., if a magnitude exceeded $15$ we took
the star as not detected in the corresponding band).  We then applied
the \textsc{Nicer} algorithm to these data, thus obtaining, for each
star, its measured extinction and related error.  Finally, we produced
smooth extinction maps by interpolating the individual extinction
measurements with three different techniques: (1) simple moving
average using Eq.~\eqref{eq:1}; (2) modified moving average using
\textsc{Nicest} [Eq.~\eqref{eq:34}]; (3) nearest neighbour
interpolation with a single star ($N = 1$).

We repeated the whole procedure $1\,000$ times, using each time a
different set of random background stars, and took the average maps
$\hat A(\vec x)$ obtained with the three techniques.  We then compared
these average maps with similar maps obtained from uniformly
distributed stars (in order to produce such maps, we applied the
completeness limit to un-extincted magnitudes).  Figures~\ref{fig:5}
and \ref{fig:6} show the results obtained, which are in clear favour
of \textsc{Nicest}.  In particular, both the simple moving average and
the nearest neighbor suffer from significant biases, up to $A_K \simeq
0.1 \mbox{ mag}$ or above, particularly in the most dense regions of
the Pipe nebula.  As shown in various papers
\citep[e.g.][]{1994ApJ...429..694L, 2006A&A...454..781L, Lombardi08a},
the amount of small scale structure present in dark molecular clouds
increases with the column density, and thus it is not surprising that
the larger biases are observed there.  In contrast, the method
presented in this paper reduces the bias by $\sim 10$, a factor
comparable to the weight number of stars $\mathcal{N}$ (which in our
simulations is $\mathcal{N} \simeq 10$ in the higher column density
regions).

The simulation performed here allows us to evaluate also the average
quadratic difference between the extinction map obtained for the
various methods $\hat A(\vec x)$ and the true (smoothed) map $\bar
A(\vec x)$, i.e.\ the quantity
\begin{equation}
  \label{eq:39}
  \bigl\langle \bigl[ \hat A(\vec x) - \bar A(\vec x) \bigr]^2
  \bigr\rangle = \mathrm{Var}\bigl[ \hat A(\vec x) \bigr] + 
  \bigl[ \bigl\langle \hat A(\vec x) - \bar A(\vec x) x \bigr\rangle
  \bigr]^2 \; .
\end{equation}
As shown by the above equation, the quantity considered here can be
written as the sum of the variance and the square of the bias of the
extinction map.  Our simulations show that, although \textsc{Nicest},
as expected, has a slightly larger variance then \textsc{Nicer}, there
is still a gain by a factor at least two in the squared difference of
Eq.~\eqref{eq:39}.  In other words, the significant reduction in the
bias performed by \textsc{Nicest} largely compensate the increase in
the scatter introduced by this method.  A naive comparison between the
results obtained from \textsc{Nicest} and from the Voronoi method
would provide even larger differences in favour of \textsc{Nicest},
but we stress that since the Voronoi method interpolates the
extinctions over a fixed number of stars (typically smaller than the
one employed by \textsc{Nicest} in our simulations) a direct
comparison is not possible.

Figure~\ref{fig:7} shows the result of simulation carried out with
the presence of foreground stars.  In particular, we generated stars
following the same prescriptions described in the previous paragraph,
but allowing for a fraction $f_0 = 0.05$ of foreground objects. 
Because of the effects of extinction, the effective fraction of
foreground objects increases in the most dense regions of the cloud,
which usually are also the ones with the larger substructures: the two
biases thus add up and can be particularly severe.  As shown by
Fig.~\ref{fig:7}, \textsc{Nicest} is effectively able to cope with
relatively large fractions of foreground stars while still providing a
virtually unbiased column density estimate.

\section{A sample application}
\label{sec:sample-application}

\begin{figure*}[!tbp]
  \begin{center}
    \resizebox{\hsize}{!}{\includegraphics[bb=52 80 567 430]{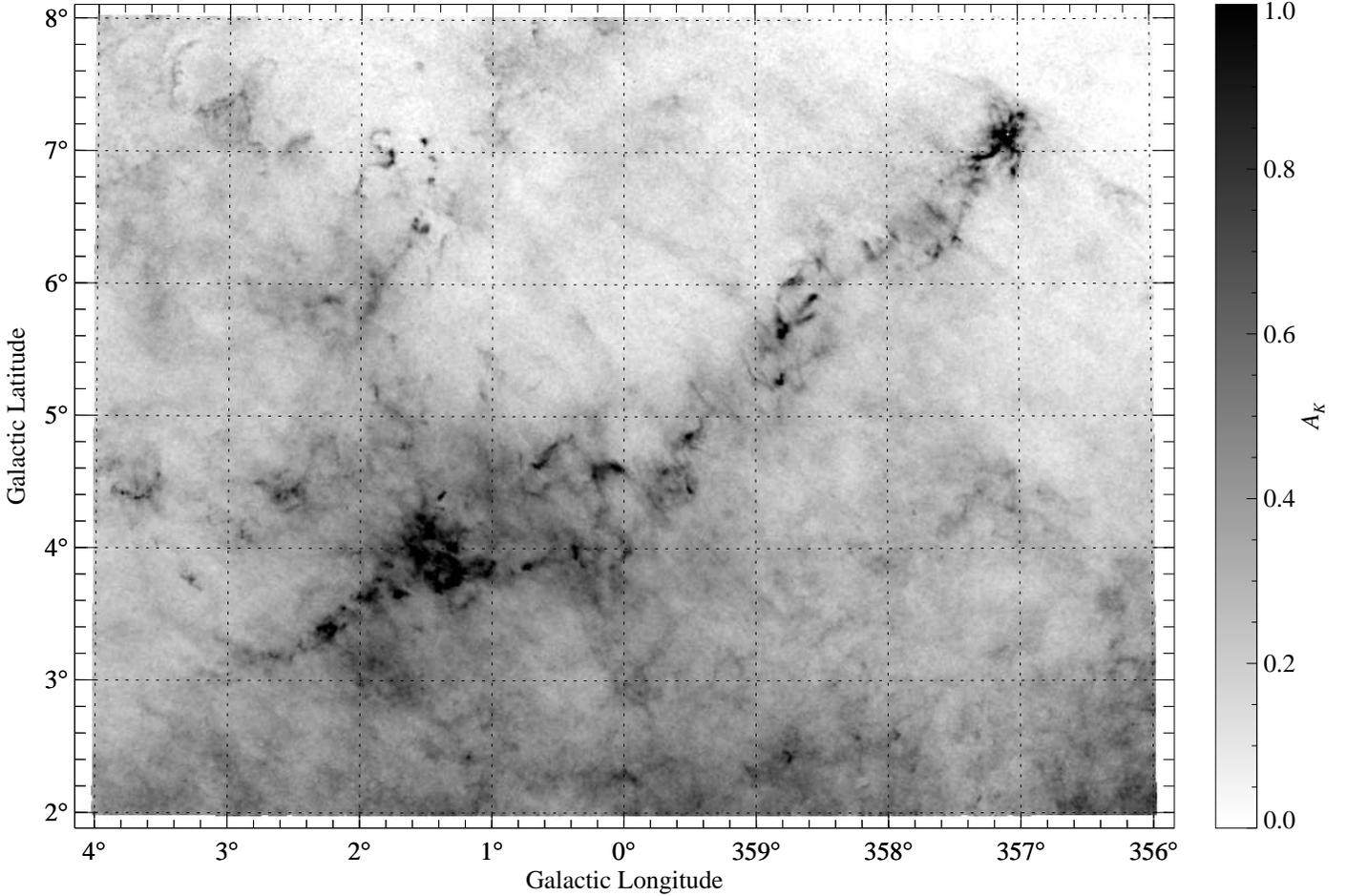}}
    \caption{The \textsc{Nicest} extinction map of the Pipe nebula,
      using the modified estimator \eqref{eq:28}.}
    \label{fig:8}
  \end{center}
\end{figure*}

The method presented in this paper was finally applied to the whole
2MASS point source catalog for the Pipe nebula region.  We used the
4.5 million stars of the 2MASS catalog located in the window
\begin{align}
  \label{eq:40}
  -4^\circ <{} & l < 4^\circ \; , & +2^\circ <{} & b < +8^\circ \; .
\end{align}
The analysis was carried out following the prescriptions of
\citet{2001A&A...377.1023L}, but using the modified estimator
\eqref{eq:34} to evaluate $\hat A$.  In particular, we generated the
final extinction map, shown in Fig.~\ref{fig:8}, on a grid of about
$1000 \times 750$ points, with scale $30 \mbox{ arcsec}$ per pixel,
and with Gaussian smoothing characterized by $\mbox{FWHM} = 1 \mbox{
  arcmin}$.  The slope of the number counts was estimated to be
$\alpha = 0.32 \pm 0.02$ in the $H$ band.

\begin{figure}[!tbp]
  \begin{center}
    \resizebox{\hsize}{!}{\includegraphics[bb=155 290 453 534]{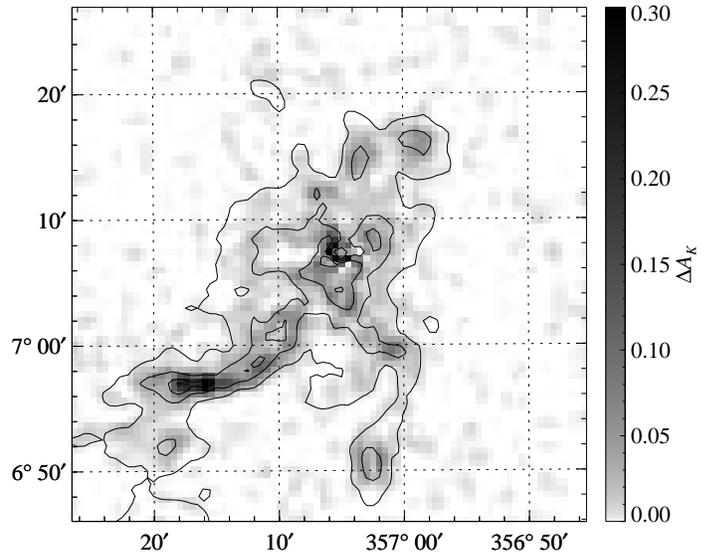}}
    \caption{The difference between the modified extinction estimates
      [Eq.~\eqref{eq:28}] and the standard ones [Eq.~\eqref{eq:1}] around
      Barnard~59.  The contour levels are at $A = \{ 0.5, 1.0, 1.5, 2.0
      \} \mbox{ mag}$ of the map of Fig.~\ref{fig:8}.}
    \label{fig:9}
  \end{center}
\end{figure}

\begin{figure}[!tbp]
  \begin{center}
    \resizebox{\hsize}{!}{\includegraphics[bb=155 290 453 534]{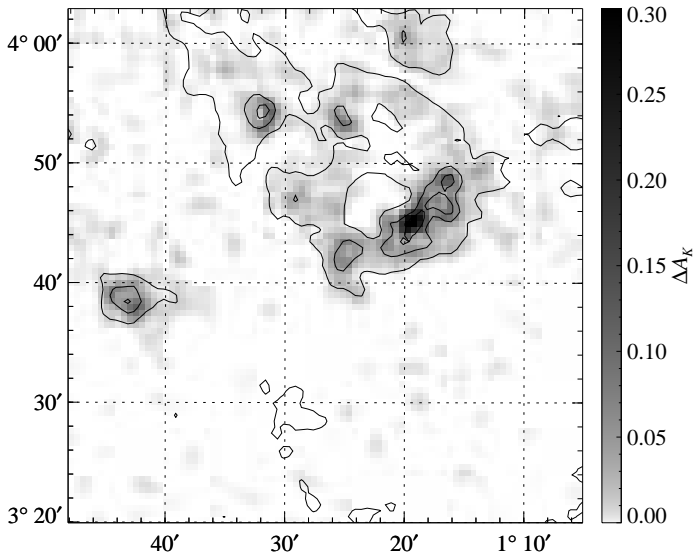}}
    \caption{The difference between the modified extinction estimates
      [Eq.~\eqref{eq:28}] and the standard ones [Eq.~\eqref{eq:1}]
      around the peak ID~3 of \citet{2006A&A...454..781L}.  The
      contour levels are at $A = \{ 0.5, 1.0, 1.5, 2.0 \} \mbox{ mag}$
      of the map of Fig.~\ref{fig:8}.}
    \label{fig:10}
  \end{center}
\end{figure}

The final, \textit{effective\/} density of stars of about $8$ stars
per pixel guarantees that the approximation used to derive the
unbiased estimator \eqref{eq:34} is valid, and that a significant
improvement over the standard \textsc{Nicer} method can be expected.
The largest extinction was measured close to Barnard~59, where $\hat A
\simeq 2.68 \mbox{ mag}$ (a value that is $0.41 \mbox{ mag}$ larger
than what obtained in \citep{2006A&A...454..781L}).

\begin{figure}[!tbp]
  \begin{center}
    \resizebox{\hsize}{!}{\includegraphics[bb=148 327 451 503]{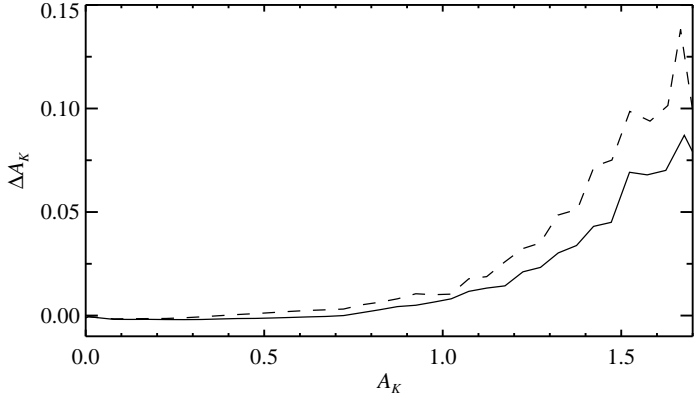}}
    \caption{The average difference $\Delta A$ between the estimators
      \eqref{eq:28} and \eqref{eq:1} as a function of the estimated
      column density $A$. The lower, solid line is relative to the
      Pipe nebula map at $\mbox{FWHM} = 1 \mbox{ arcmin}$ resolution; 
      the upper dashed line is at for a resolution $\mbox{FWHM} = 1.5
      \mbox{ arcmin}$.  The latter plot is essentially what we would
      obtain at a $\mbox{FWHM} = 1 \mbox{ arcmin}$ if the distance of
      the Pipe nebula increases by a factor $1.5$.}
    \label{fig:11}
  \end{center}
\end{figure}

Figures~\ref{fig:9} and \ref{fig:10} show a comparison between dense
regions mapped using \textsc{Nicer} and \textsc{Nicest}.  We note
that, as expected, the two methods are equivalent in low-density
regions, while the new one consistently estimates larger column
densities as $A$ increases.  The same effect can be appreciated more
quantitatively from Fig.~\ref{fig:11}, where we plot the relationship
between the average estimates $\hat A$ obtained in
\citet{2006A&A...454..781L} and here.

Finally, we argue that the plot of Fig.~\ref{fig:11} is strongly
related to Fig.~9 of \citet{2006A&A...454..781L} where we show the
increase in the scatter of $\hat A_n$ for the different stars used in
each point of the extinction map as a function of the average $\hat A$
in the point.  This $A$-$\sigma$ relation is most likely due to
unresolved substructures in the cloud, that are expected to be more
prominent in high-density regions.  Recently, we reconsidered the
$A$-$\sigma$ relation and defined a quantity called $\Delta^2(\vec x)$
\citep{Lombardi08a}.  This quantity is simply related to the local
scatter of measured extinctions, but also properly takes into account
the contribution of measurement errors to the observed scatter.
Interestingly, this quantity can be defined in terms of simple
observables, and can be shown to be directly related to the local
scatter of column densities:
\begin{equation}
  \label{eq:41}
  \Delta^2(\vec x) = \frac{\sum_n w_n \Delta_n^2}{\sum_n w_n} \; ,
\end{equation}
where $\Delta_n \equiv A(\vec x_n) - \bar A(\vec x)$ is the difference
between the column extinction at the position of the $n$-th star and
the local average column extinction.  A comparison of
Eq.~\eqref{eq:41} with Eq.~\eqref{eq:8} clearly shows that, except for
a numerical factor $\beta$, the bias expected in the standard
\textsc{Nicer} method is simply proportional to the $\Delta^2(\vec x)$
of \citep{Lombardi08a}.

\section{Discussion}
\label{sec:discussion}

The numerical simulations and a first sample application of
\textsc{Nicest} have shown that the method presented in this paper can
significantly alleviate the bias introduced by small-scale structures
and by foreground stars in extinction studies.  Of course,
\textsc{Nicest} too has some limitations, which however are largely
unavoidable (and thus inherent to any extinction-based method).

First, we note that in order for \textsc{Nicest} to work effectively,
the weight number of background stars $\mathcal{N}$ must be
significantly larger than unity: as shown in
Sect.~\ref{sec:fix-inhom-bias}, $\mathcal{N}$ is directly related to
the reduction in the bias provided by \textsc{Nicest} with respect to
\textsc{Nicer}, and thus having $\mathcal{N} \sim 1$ does not give any
benefit.  This point is also important when correcting for foreground
star contamination.  For example, if $\mathcal{N} \sim 1$ and the
local fraction $f$ of foreground stars is large (e.g., because we are
in a particularly dense core), on average we do not expect any
background star, and we will thus consistently measure a vanishing
extinction.  Hence, \textsc{Nicest}, like any other extinction based
method, only works if there are a sizable number of background stars
that can be used for reliable extinction measurements.

The above point might be interpreted as an exceedingly stringent
requirement for the smoothing window used in \textsc{Nicest}.  In
fact, if a weight function $w(\vec x'; \vec x) = w(\vec x' - \vec x)$
invariant upon translation is used, this function should be taken
broad enough to guarantee that the weight number of background stars
$\mathcal{N}(\vec x)$ is significantly larger than unit everywhere.
This, in turn, would imply that $\mathcal{N} \gg 1$ in the
intermediate extinction regions, i.e.\ that the extinction map in
these regions has a poor resolution, well below the limits imposed by
the density of background stars.  In reality, the whole derivation of
the statistical properties of the \textsc{Nicest} technique is still
valid for weight functions which are not spatially invariant.  Hence,
one does not need to use a fixed window size for $w(\vec x'; \vec x)$,
but rather this function could be taken to change shape for different
locations $\vec x$.  For example, a simple scheme could be the use of
a Gaussian shape for $w(\vec x'; \vec x)$ with the typical scale
chosen according to a local estimate of the density of background
stars, in a way such that $\mathcal{N}(\vec x) \sim \mbox{const} \sim
10$.  This choice would guarantee an optimal resolution everywhere,
and would still allow one to make use of the \textsc{Nicer} technique
to reduce the inhomogeneity bias.

Another point to keep in mind is that \textsc{Nicest} gives a large
weight to red sources.  This has two potential unwanted consequences.
First, it introduces a small bias, that has been corrected to first
order in Eq.~\eqref{eq:34}.  Second, as shown also by the numerical
simulations described in Sect.~\ref{sec:sample-application}, it
slightly increases the noise in the final map.  Again, since the noise
in final map $\hat A(\vec x)$ is proportional to $1 / \mathcal{N}$,
the solution is to make sure that $\mathcal{N}$ is sufficiently large.
We believe that a small increase in the noise is a fair price to pay
for the significant reduction in the bias provided by \textsc{Nicest}
over \textsc{Nicer} (see also \citealp{Eadie}).

A potentially more severe problem can be young stellar objects (YSOs)
with infrared excess.  These sources can be present in the most dense
star-forming cores of molecular clouds, where they can severely bias
our method.  We note, however, that since these sources have peculiar
colors and are not usually present in the control field used for the
calibration, they represent anyway a problem for all extinction based
methods, including \textsc{Nicer} and the nearest neighbours ones.  A
median estimator is in principle safer to use in these cases, as long
as the YSOs present in the core are a minority of the total number of
background stars observed in the region; unfortunately, the tendency
of YSOs to appear in clusters does not help here (note also that a
median is usually more noisy than the simple average).  In any case,
the obvious solution is thus to exclude as much as possible YSOs from
the list of sources used in the extinction maps.

Finally, let us briefly mention alternative possibilities to reduce
the bias considered in this paper.  Since the substructure bias is due
to a relationship between the local density $\rho(\vec x)$ and the
extinction map $A(\vec x)$ [Eq.~\eqref{eq:2}], one could try to use a
tracer of the local density to perform the correction.  We discussed
already a method based on this concept, the nearest neighbours
interpolation.  However, as shown in Sect.~\ref{sec:simulations}, this
method fails (see also Sect.~\ref{sec:nearest-neighbors}).  In
general, a problem with methods based on the local estimate of the
density of background stars is that one needs to be able to obtain
$\rho(\vec x)$ on scales smaller than the ones that characterize the
weight $w(\vec x'; \vec x)$, or otherwise it is not possible to give
different weights to the different stars that contribute significantly
to the weighted average of Eq.~\eqref{eq:1}.  However, if the density
of background stars is large enough to have a good handle on the local
density $\rho(\vec x)$, one basically does not need to care about
substructures, because it is always possible to make extinction maps
at the same resolution as the density map.  Hence, the only effective
way to deal with substructures involves a use of the local extinction
measurements, as done in \textsc{Nicest}.  Note also that any
density-based correction will be completely ineffective with
foreground stars, in contrast to the method presented here.

\section{Conclusions}
\label{sec:conclusions}

The main results obtained in this paper can be summarized in the
following points:
\begin{itemize}
\item We discussed the effects of small scale inhomogeneities in the
  NIR extinction maps based on color excess methods, and showed that
  large inhomogeneities can significantly bias standard extinction
  maps toward low column densities.
\item We proposed a new estimator for $\hat A$, \textsc{Nicest}, and
  we showed that it is (i) equivalent to the usual estimator in
  low-density regions, (ii) \textit{unbiased\/} and \textit{robust\/}
  for any value of $A$, i.e.\ insensitive to the actual form and
  amount of substructure present in the cloud.
\item We showed that the new estimator is also suitable in presence
  contamination by foreground stars.
\item We tested \textsc{Nicest} against numerical simulations, and
  showed that it effectively reduces by a large factor the biases due
  to substructures and to foreground stars.  We also tested an
  alternative approach, the nearest neighbor, and showed that the
  results obtained from this interpolation are still severely biased.
\item We applied \textsc{Nicest} to the Pipe nebula and showed a few
  preliminary properties of the resulting extinction map.  We also
  noted a direct connection between the bias of the \textsc{Nicer} and
  the $\Delta^2$ map defined by \citep{Lombardi08a}.
\end{itemize}

\acknowledgements 

We thank J.~Alves, C.J.~Lada, and G.~Bertin for stimulating
discussions and suggestions, and the referee, L. Cambresy, for helping
us improve the paper.  This research has made use of the 2MASS
archive, provided by NASA/IPAC Infrared Science Archive, which is
operated by the Jet Propulsion Laboratory, California Institute of
Technology, under contract with the National Aeronautics and Space
Administration.

\bibliographystyle{aa} 
\bibliography{../dark-refs.bib}

\end{document}